# In-vivo imaging of the human thalamus: a comprehensive evaluation of structural magnetic resonance imaging approaches for thalamic nuclei differentiation at 7T

Cristina Sainz Martinez,[1,2] José P. Marques,[3] Gabriele Bonanno,[4,5,6] Tom Hilbert,[4,7,8] Constantin Tuleasca,[9,8,10] Meritxell Bach Cuadra,[2,7] João Jorge[1,*]

1. CSEM - Swiss Center for Electronics and Microtechnology, Switzerland;

2. CIBM Center for Biomedical Imaging, Switzerland;

3. Donders Centre for Cognitive Neuroimaging, Radboud University, Nijmegen, Netherlands;

4. Advanced Clinical Imaging Technology, Siemens Healthineers International AG, Bern, Switzerland;

5. Translational Imaging Center (TIC), Swiss Institute for Translational and Entrepreneurial Medicine (SITEM), Bern, Switzerland;

6. Departments of Radiology and Biomedical Research, University of Bern, Switzerland;

7. Department of Radiology, Lausanne University Hospital (CHUV) and University of Lausanne (UNIL), Lausanne, Switzerland;

8. Signal Processing Laboratory 5 (LTS5), École Polytechnique Fédérale de Lausanne, Lausanne, Switzerland;

9. Department of Clinical Neurosciences, Neurosurgery Service and Gamma Knife Center, Centre Hospitalier Universitaire Vaudois (CHUV), Lausanne, Switzerland

10. Faculty of Biology and Medicine, University of Lausanne (UNIL), Lausanne, Switzerland

## Abstract

The thalamus is a subcortical structure of central importance to brain function, which is organized in smaller nuclei with specialized roles. Despite significant functional and clinical relevance, locating and distinguishing the different thalamic nuclei in vivo, non-invasively, has proved challenging with conventional imaging techniques, such as $T_1$ and $T_2$-weighted magnetic resonance imaging (MRI). This key limitation has prompted extensive research efforts, and several new candidate MRI sequences for thalamic imaging have been proposed, especially at 7T. However, studies to date have mainly been centered on individual techniques, and often focused on subsets of specific nuclei. It is now critical to evaluate which options are best for which nuclei, and which are globally the most informative. This work addresses these questions through a comprehensive evaluation of thalamic structural imaging techniques in humans at 7T, including several variants of $T_1$, $T_2$, $T_2^*$ and magnetic susceptibility-based contrasts. All images were obtained from the same participants, to allow direct comparisons without anatomical variability confounds. The different contrasts were qualitatively and quantitatively analyzed with dedicated approaches, referenced to well-established thalamic atlases. Overall, the analyses showed that quantitative susceptibility mapping (QSM) and $T_1$-weighted MP2RAGE tuned to maximize gray-to-white matter contrast are currently the most valuable options. The two contrasts display unique, complementary features and, together, enable the distinction of the majority of known nuclei. Likewise, their combined information could provide a powerful input for automatic segmentation approaches. To our knowledge, this study represents the most comprehensive assessment of structural MRI contrasts for thalamic imaging to date.

**\*Corresponding author:**

João Jorge, PhD

CSEM
Freiburgstrasse 2
3008 Bern, Switzerland

e-mail: joao.jorge@csem.ch

# 1. Introduction

The thalamus is a deep brain structure of central importance, which acts as a gateway for sensory and motor signals to the cerebral cortex (Sherman and Guillery 2006). It is organized in smaller specialized nuclei with different key functions, such as relaying and modulating information for sensory modalities, motor control, and regulation of sleep, consciousness and emotions (Saalmann and Kastner 2011; Gent et al. 2018). Various neurological and psychiatric disorders have been associated with alterations in thalamic structure and function. For instance, studies have reported reduced thalamic volume in patients with multiple sclerosis (Bergsland et al. 2021), schizophrenia (Byne et al. 2009), Alzheimer's disease (Pardilla-Delgado et al. 2021), bipolar disorder (Radenbach et al. 2010) and major depressive disorder (Nugent et al. 2013). Changes in functional connectivity between thalamic and other brain regions have been observed in conditions such as epilepsy, Parkinson's disease and schizophrenia (Woodward, Karbasforoushan, and Heckers 2012). These effects could lead to the identification of promising biomarkers for the respective pathologies. At the same time, several nuclei have also been established as effective targets for clinical interventions, involving activity modulation with deep brain stimulation (DBS), focused irradiation with stereotactic radiosurgery (SRS) (Kooshkabadi et al. 2013; Witjas et al. 2015) or controlled thermocoagulation, as in high-intensity focused ultrasound (HIFU) techniques (Elias et al. 2016; Rohani and Fasano 2017). For example, DBS of the anterior nucleus (AN) has been shown to reduce seizure frequency and improve cognitive function in patients with refractory epilepsy (Kerrigan et al. 2004; Fisher et al. 2010); targeting the ventral intermediate nucleus (Vim) by HIFU has been effective in reducing the motor effects of essential tremor (Elias et al. 2016).

Minimally-invasive interventional techniques such as SRS and HIFU, and non-invasive analysis techniques such as image-based morphometry, require accurate localization and delineation of the thalamic nuclei, which is currently challenging to obtain non-invasively. Despite its outstanding soft-tissue contrast, magnetic resonance imaging (MRI) based on conventional contrasts such as $T_1$ and $T_2$ weighting yields little to no distinguishable features in the thalamus (Boutet et al. 2021). As a result, minimally invasive thalamotomies (particularly by SRS) must rely on indirect targeting methods using nearby anatomical landmarks (such as the anterior and posterior commissure, thalamic height and laterality to the third ventricular wall) (Witjas et al. 2015). Naturally, indirect targeting techniques cannot fully account for fine-scale individual anatomical variability, nor structural deviations due to age and disease (Mai and Forutan 2012), which may therefore variably impact the effectiveness of the interventions (Ohye et al. 2012). Additionally, in SRS, the clinical effect (e.g. tremor arrest) is not immediate, and hence does not allow intraoperative confirmation of the targeting accuracy (Tuleasca et al. 2018; Niranjan et al. 2017).

The current gold standards for thalamic nuclei localization are post-mortem histology and in-vivo electrophysiology with depth electrodes during DBS. Histological stains can depict different nuclei structures and composition depending on the staining technique, and have enabled the creation of useful thalamic atlases such as the so-called Schaltenbrand atlas (Schaltenbrand and Wahren 1977), based on silver staining of myelin, and the Morel atlas (Morel, Magnin, and Jeanmonod 1997), built with multiple stainings. These standard atlases have proved highly valuable as references for in-vivo research, but are likewise challenging to align precisely with in-vivo images, due to anatomical variability and the afore-mentioned lack of image contrast in the thalamus. Electrophysiology, on the other hand, allows direct measurements of local neuronal activity in living individuals, and thereby a relatively precise localization of nuclei of interest based on their functional signature (Albe-Fessard et al. 1966).

However, the need to implant depth electrodes makes the approach more invasive, and limited in field-of-view (FOV) to relatively localized regions in each implantation. Thus, altogether, these gold standard methods carry strong constraints in terms of feasibility, invasiveness, and coverage of the targeted area (Eisinger et al. 2018).

Given the key functional and clinical relevance of the thalamus, and the limitations of standard imaging modalities, extensive research efforts have been oriented in recent years to the development of more effective imaging techniques to differentiate the thalamic nuclei non-invasively, especially with MRI. The explored approaches include structural, diffusion and functional modalities (Iglehart et al. 2020). Diffusion studies have managed to distinguish certain nuclei groups based on their local diffusion characteristics (Battistella et al. 2017; Najdenovska et al. 2018), or their structural connectivity with different cortical regions (Behrens et al. 2003). With fMRI, typically in resting state (Bolton et al. 2024; Tuleasca et al. 2019), the voxels in the thalamus can be clustered based on the similarity of their signal timecourses, or on their similarity to different cortical regions (Hale et al. 2015; Kim, Park, and Park 2013; Yuan et al. 2016; Kumar et al. 2017). Additionally, task paradigms, such as motor tasks and sensory stimulation, have been explored to identify nuclei involved in these processes (Charyasz et al. 2023). Despite some success, both diffusion and functional approaches do exhibit limitations in terms of spatial specificity, typically identifying mainly larger nuclei and nuclei groups. The time required for acquisition and for the analyses can also make them less practical. Additionally, fMRI results can suffer from substantial inter-subject and inter-session variability, typical of undirected resting-state network activity (Specht 2020).

Notably, structural MRI approaches can typically achieve higher spatial resolutions than diffusion and fMRI, and are faster to acquire and to process – making them more suitable for clinical applications. Earlier works explored approaches based on $T_1$ weighting, namely with images acquired at the gray matter (GM)- (Magnotta et al. 2000; Bender et al. 2011) and white matter (WM)-nulling point (Sudhyadhom et al. 2009) of the inversion recovery (IR). These selective suppressions enabled distinctions between more medial and more lateral nuclei. More recently, studies at ultra-high field (7T) have opened unprecedented possibilities for thalamic imaging: on the one hand, the higher signal-to-noise ratio (SNR) at 7T enables reliable acquisitions at sub-millimeter spatial resolution, which is highly advantageous to image small nuclei like the Vim. On the other hand, the increased field strength brings additional advantages to contrasts based on $T_1$ (longer relaxation times) and crucially to magnetic susceptibility (stronger effects), which have been found promising for thalamic visualization. In particular, $T_1$-based approaches relying on WM suppression at 7T have been found superior to their equivalents at 3T (Saranathan et al. 2021; 2015), and promising observations have been reported with the MP2RAGE approach (Marques et al. 2010), when optimized specifically to maximize GM-to-WM contrast (Marques and Gruetter 2013). IR-turbo spin echo (IR-TSE) has also been reported to reveal specific thalamic features in anterior regions, particularly thanks to the $T_1$-weighting (Kanowski et al. 2014). Regarding susceptibility, initial studies found that susceptibility weighting imaging (SWI) (Haacke et al. 2009), a technique originally developed to enhance venous vessels, can reveal exquisite features in the thalamic tissues as well, with solid agreement with standard atlas information (Abosch et al. 2010). Subsequent work has shown that the phase-magnitude combination process used in SWI can be tuned to further enhance intra-thalamic contrast, greatly accentuating these features (Jorge et al. 2020). Other studies have explored the use of quantitative susceptibility mapping (QSM), which provides direct estimates of the underlying property of magnetic susceptibility (Ruetten, Gillard, and Graves 2019). QSM likewise shows highly promising features in the thalamus (Deistung et al. 2013), consistent with SWI and atlas information, thereby supporting that magnetic susceptibility does indeed vary meaningfully within the thalamic anatomy.

Altogether, the current body of literature on thalamic structural imaging at 7T indicates that there are valuable and effective options to image the thalamus. However, each of the existing studies to date has typically focused on one specific technique, and often directed its attention to specific thalamic nuclei. Important questions are currently unanswered: (i) What is the best technique available to image specific thalamic nuclei? (ii) If no specific nucleus is targeted a priori (e.g., for connectivity studies), which techniques can be used to distinguish the largest diversity of thalamic nuclei? (iii) Which nuclei remain challenging to delineate based on structural MR imaging?

To address these questions, in the present work, we have conducted a comprehensive and systematic evaluation of thalamic structural imaging techniques for humans at 7T. We implemented the most relevant sequences based on $T_1$, $T_2$ and $T_2$*, as well as magnetic susceptibility. Importantly, all sequences were acquired from the same participants, to allow direct comparisons across sequences without anatomical variability confounds. The different contrasts were qualitatively and quantitatively analyzed, using dedicated methods for spatial alignment and to assess nuclei distinguishability, with well-established thalamic atlases as support. This allowed assessing the degree to which each sequence can differentiate each thalamic nucleus, and therein clarify the questions defined above.

## 2. Methods

This study was approved by the local ethics committee (KEK Bern) and involved the participation of 4 healthy adult volunteers (2 male/2 female, 25–30 years old), who provided written informed consent. All image and atlas processing was performed in Python, combined with tools from SigPy (Ong and Lustig, n.d.), ANTs (Avants et al. 2008) and ITK-SNAP (Yushkevich, Gao, and Gerig 2016).

### 2.1 Image acquisition and reconstruction

The acquisitions were performed at 7T (MAGNETOM Terra, Siemens Healthineers AG, Erlangen, Germany) using a single-channel transmit, 32-channel receive head RF coil (Nova Medical, Wilmington MA, USA). The following data were obtained from each subject (with sequence parameters detailed in Table *1*):

1. $B_1^+$ mapping using a 3D SA2RAGE sequence (Eggenschwiler et al. 2011). The acquisition was reconstructed online to yield a $B_1^+$ map.

2. Compressed sensing (CS)-accelerated 3D MP2RAGE (Mussard et al. 2020). Reconstructed online to yield a conventional $T_1$-weighted image (hereafter labeled $T_1$w) and a $T_1$ map.

3. 3D MPRAGE tuned to suppress WM signal (Tourdias et al. 2014). Reconstructed online to yield a WM-nulled image ($WM_{null}$).

4. CS-accelerated 3D MP2RAGE tuned to maximize GM-to-WM contrast (Marques and Gruetter 2013). Reconstructed online to yield a contrast-optimized $T_1$-weighted image ($GWM_{opt}$).

5. Flow-compensated 3D multi-echo GRE (ME-GRE). To allow using the COSMOS method (T. Liu et al. 2009), the acquisition was repeated for different head orientations (9, 5, 6 and 6 for subjects 1–4, respectively, depending on session time constraints). The raw data were reconstructed offline using Python tools developed in-house together with SigPy: first, the five $T_2$*-weighted complex images were reconstructed using SENSE with wavelet-based regularization, employing complex coil sensitivity maps estimated from the data using ESPIRiT (Uecker et al. 2014), combined with a virtual "body coil-like" channel estimated via block coil compression, to serve as a singularity-free reference (Bilgic et al. 2016; 2017). From the reconstructed magnitude images (first orientation acquisition), an $R_2^*$ map was estimated with a weighted least-squares fit of mono-exponential $T_2^*$ relaxation across the echoes. Field maps were estimated from the phase data of each orientation, and underwent background field removal using V-SHARP (W. Li et al. 2014). QSM was then estimated using two distinct approaches: (i) COSMOS ($QSM_{COS}$), combining the field maps from all orientations (T. Liu et al. 2009), and (ii) STAR-QSM ($QSM_{STAR}$), using only the first orientation (Wei et al. 2015). Both methods were employed in order to investigate, in parallel, the value of magnetic susceptibility for thalamic imaging in general (based on COSMOS, considered a gold-standard for dipole inversion), and the value of QSM acquisitions in practice (based on $QSM_{STAR}$, which is more routinely feasible). Both methods were previously implemented in Python and validated using the BigBrain-MR computational phantom (Sainz Martinez, Bach Cuadra, and Jorge 2023). For COSMOS, the different orientation images were first aligned by linear registration using ANTs.

6. Flow-compensated 3D GRE sequence. The raw data were reconstructed offline similarly to the ME-GRE data to yield a magnitude and a phase image; an SWI-contrast image (Haacke et al. 2009) was then generated using the optimized phase-magnitude combination proposed in (Jorge et al. 2020).

7. Variable flip-angle (VFA) 3D TSE SPACE (Mugler 2014), reconstructed online to yield a conventional $T_2$-weighted image ($T_2$w).

8. IR-prepared VFA 3D TSE SPACE adapted to approximate (Kanowski et al. 2014), reconstructed online to yield an IR-prepared TSE image (IRTSE).

Alongside the sequence parameters detailed in Table 1, all images were acquired with a whole-brain FOV, except for SWI and IRTSE, which covered a smaller slab centered on the thalamus. The $B_1^+$ map was acquired first, and used to calibrate the transmit voltage specifically for the central region where the thalamus is located – which typically showed $B_1^+$ values 10–20% higher than the brain average.

## 2.2 Atlas registration

The thalamic structural features obtained with the different contrasts were analyzed with respect to two standard atlases published from histological studies. We did not consider atlases generated from MRI data (e.g. the recent, finely-detailed MuSus atlas, built from $T_1$w and QSM (He et al. 2022)), to avoid inherent biases towards specific contrasts in the present comparisons.

A multi-step registration procedure was implemented to align the acquired images with the atlases (*Fig. 1*a). First, each subject's images were all aligned to the $WM_{null}$ contrast, using ANTs with affine registration and mutual information as optimization metric. One of the adopted atlases was a simplified version of Morel (Morel, Magnin, and Jeanmonod 1997) developed by (Su et al. 2019); it was non-linearly aligned to the $WM_{null}$ image of each subject using the THOMAS method, developed by the same authors for this purpose, using non-linear Symmetric Normalization ("SyN", comprising affine + deformable transformations) from ANTs. The subsequent analyses based on this atlas were performed directly in each individual's $WM_{null}$ native space. The second atlas was from (Schaltenbrand and Wahren 1977), in MNI space. Since this atlas is not available in 3D digital form, we focused specifically on an axial slice at $Z = 3.5$mm, which contains the Vim and diverse other nuclei (Abosch et al. 2010); the individual subject data were warped to MNI space via the WM$_{null}$, using the ICBM T$_1$w image as reference (ICBM 152, version 2009c, http://nist.mni.mcgill.ca/atlases/), non-linearly with ANTs (SyN).

[see *Fig. 1* on page 22]

## 2.3 Contrast analysis

In pursuing an objective comparison between the features observed in the acquired images and the nuclei organization defined by the histological atlases, we identified two important challenges:

i. **Anatomical misalignment:** Even in images where a nucleus is well distinguishable, anatomical deviations with respect to the standard atlases are still expected to occur due to natural individual variability (Mai and Forutan 2012), and to registration imperfections, which have been observed even with dedicated approaches such as THOMAS (Williams, Roesch, and Christakou 2022);

ii. **Complex contrast features:** In most cases where nuclei can be distinguished on the acquired contrasts, this distinction is based on general intensity differences between the regions of interest (ROIs) formed by those nuclei; however, in some cases, the distinction is instead possible based on *contours* observed delineating the nuclei (e.g., see QSM in Fig. *3*), without necessarily showing strong differences in intensity inside the ROIs.

Thus, for instance, a comparison approach based on estimating intensity differences or contrast-to-noise ratio (CNR) between atlas-based ROIs defining each nucleus can be limited and biased by both problems mentioned above – anatomical deviations with respect to the atlas ROIs, and distinction not based on average intensities. Instead, for this study, we designed a qualitative and a quantitative approach specifically tailored to accommodate these challenges, as described below.

### 2.3.1 Qualitative analysis

A preliminary visual assessment of the images was first conducted to review both the general quality of each acquisition and its consistency across subjects. Subsequently, a more systematic analysis was carried out to compare the effectiveness of each contrast in distinguishing each pair of thalamic nuclei (*Fig. 1*b). The analysis was performed independently by two investigators with extensive experience in thalamic MRI (co-authors CSM and JJ, hereafter named "evaluators"), as follows: each evaluator considered every pair of neighboring nuclei as defined by the Morel atlas, and for each pair indicated at most the two contrasts that allowed the best visual distinction between the two nuclei (either based on intensity difference or on a clear border/contour), across all subjects (or none if no contrasts allowed that distinction); the independent assessments of the two evaluators were then merged into one by "overlap", i.e. by selecting the contrasts indicated by both evaluators for each nuclei pair. The analysis was also repeated using the Schaltenbrand atlas (slice of interest) as reference for the nuclei definition.

### 2.3.2 Quantitative analysis

This approach sought to complement the qualitative analysis by quantifying how effectively each image can distinguish each pair of adjacent thalamic nuclei, in terms of a change in inner intensity or a well-defined border (*Fig. 1*c). The basic idea of the approach was to sample the image intensities along a curve going from the center of a nucleus to the center of its neighbor, and then evaluating how well this profile matches a step-like function (for cases where the nuclei have different overall intensities) and/or a central "peak" or "valley" (for cases differentiated by their border contour). To facilitate visual checks and ensure the correctness of the trajectories and extracted profiles, we focused solely on the Schaltenbrand atlas at slice $Z = 3.5$ mm for nuclei definition, and therefore worked in 2D. Specifically, for each nucleus, a point corresponding approximately to its geometric center was defined (with manual adjustments where needed, to accommodate the intricate geometry of some of the nuclei). For each pair of adjacent nuclei, a point was also defined on the boundary between them (as delineated by the atlas), positioned approximately halfway along the length of their border. A parabolic trajectory on the 2D slice was then fitted to these three points (two nuclei centers and the border point) (*Fig. 1c*). To extract the intensity profile, the parabola was segmented in a series of equidistant points, and the image intensities at these points were sampled through linear interpolation from the image voxel grid. The mean intensity across the profile was removed, to focus on the variations.

Each extracted intensity profile $f(x)$ was then evaluated by non-linear fitting to a model of the form:

$$\hat{f}(x) = a_0 e^{\frac{-(x-a_1)^2}{2a_2^2}} + \frac{a_3}{1 + e^{-a_4(x-a_5)}} - a_6 \qquad \text{Eq. 1}$$

where $x$ is the position along the parabolic profile between the two nuclei centers, and $a_{1-6}$ are fitting parameters. The first term of the model is a gaussian function, to allow describing a border between the two nuclei ("peak"/"valley"), and the second term is a sigmoid function, to describe a "step-like" intensity change. The position and "width" of these shapes is controlled by the fitting parameters, and therefore adaptive. Altogether, these characteristics provide the necessary flexibility to overcome the previously described obstacles (i) and (ii). The fitting was performed with the function *curve_fit* from the optimization library of SciPy. Given the high flexibility of the model, boundary conditions were empirically defined for each parameter to avoid unsuitable fits that would not reflect visible nuclei differences. In particular, the centers of the sigmoid and gaussian ($a_{1,5}$) were restricted from residing at the extremes of the profile, the amplitudes ($a_{0,3}$) were limited based on the amplitude range of the measured profile, and the widths ($a_{2,4}$) were restricted with respect to the profile length.

After fitting, the coefficient of determination $R^2$ of the fit was estimated as:

$$R^2 = 1 - \frac{\sum_i f(x_i) - \hat{f}(x_i)}{\sum_i f(x_i)} \qquad \text{Eq. 2}$$

where the index $i$ iterates through all the points of the profile. $R^2$ varies between 0 (a fit that explains no variance in the data) and 1 (a fit that explains all the variance). This metric was therein adopted as a proxy for nuclei distinguishability in each image: if the intensity variation when moving from one nucleus to its neighbor is dominated by a step and/or by a peak/valley, the model is expected to explain a large part of that variance, resulting in a large $R^2$. Benefitting from the 2D restriction chosen for this analysis, we could visually review all trajectories, the corresponding profiles and model fits, to ensure an adequate behavior and meaningful results.

# 3. Results

## 3.1 Data quality

Overall, the acquisition protocol successfully produced images that were visually comparable to the respective literature sources, and thus allowed reproducing their results (*Fig. 2*, *Supp. Fig. 1*). An exception was the IRTSE contrast, originally proposed for the thalamus by (Kanowski et al. 2014), where a particularly good distinction of the anterior nuclei was reported, which we could not achieve with our protocol. Through visual inspection, most contrasts were generally consistent across subjects, in terms of contrast- and signal-to-noise, with a few noteworthy exceptions: (i) the $T_1$w contrast exhibited moderate but visible SNR variability across subjects, with subjects 2 and 4 showing higher SNR than the others; (ii) the QSM data also exhibited more variable quality – $QSM_{COS}$ was of particularly high quality for subject 1, where 9 orientations were used, compared to the others (6 or less orientations); the image quality was visibly poorer for subject 2, possibly linked to more accentuated artifacts that could be observed in the respective phase images.

[see *Fig. 2* on page 23]

## 3.2 Qualitative analysis

Overall, based on the qualitative analysis, QSM (both $QSM_{COS}$ and $QSM_{STAR}$) and $GWM_{opt}$, together, allowed the clearest distinctions *between* almost every nuclei pair considered (*Fig. 3*, Table 2, Table 3). Other $T_1$-based images ($T_1$map, $WM_{null}$) appeared sensitive to roughly the same features as $GWM_{opt}$, but with lower SNR and contrast. Regarding the susceptibility-based images, SWI, while showing a similar trend to QSM, was more sensitive to the presence of veins, which are not of interest to this application. The $R_2$*map lacked contrast within the thalamic region. Conventional contrasts such as $T_1$w and $T_2$w offered limited insights into the thalamus, as did the IRTSE approach. These findings were generally consistent across subjects (*Supp. Fig. 2*, *Supp. Fig. 3*, *Supp. Fig. 4*).

[see *Fig. 3* on page 24]

With reference to the **Morel-based atlas** (*Fig. 3* (left), Table 2), it was observed that both $QSM_{COS}$ and $QSM_{STAR}$ enabled the distinction of numerous nuclei, such as the Ventral Posterior Lateral (VPL), Ventral Lateral Posterior-inferior (VLP-inf) and Ventral Lateral anterior (VLa) nuclei, for example (Fig. 3-Slice ii). Only in the case of subject 2, whose QSM images were of lower quality, certain nuclei, particularly in inferior slices, such as the Lateral geniculate (Lgn) and Medial geniculate nuclei (Mgn), could not be as clearly distinguished (*Supp. Fig. 2*). Importantly, in superior regions (*Fig. 2*-Slice iii), the presence of large thalamic draining veins was well visible in susceptibility-based contrasts like SWI, hindering nuclei visualization. Among the $T_1$-weighted sequences, $GWM_{opt}$ consistently provided the clearest distinctions across the medial-lateral direction, outperforming $WM_{null}$, $T_1$w and $T_1$map in SNR and contrast, across all subjects. $GWM_{opt}$ proved particularly valuable in distinguishing nuclei pairs such as the Mediodorsal-Parafascicular (MD-Pf) and Ventral Lateral Posterior-superior (VLP-sup) nuclei. Still, a few superior-anterior nuclei, such as the VLP-sup, Anterior Ventral (AV) and Ventral Anterior (VA), remained challenging to differentiate with all contrasts tested (*Fig. 3*- Slice iii), and across all subjects (*Supp. Fig. 2*, *Supp. Fig. 3*, *Supp. Fig. 4*-slice iii).

When comparing the two QSM approaches, the multi-orientation QSM$_{COS}$ did consistently exhibit superior image quality compared to the single-orientation QSM$_{STAR}$, displaying a higher SNR and reduced streaking artifacts arising, for instance, from high susceptibility veins. A few nucleus borders, such as between the AV and the Mediodorsal-Parafascicular (MD-Pf) located in superior regions, proved more challenging to image with QSM$_{STAR}$ than with QSM$_{COS}$ (*Fig. 3*-slice iii). Nevertheless, QSM$_{STAR}$ was still adequate for distinguishing most pairs of nuclei that could be discerned by QSM$_{COS}$.

[see Table 2 on page 29]

With reference to the **Schaltenbrand atlas**, the results closely aligned with those described for the Morel-based atlas (*Fig. 3* (right), Table 3). SWI and, more prominently, QSM, provided the clearest and most comprehensive distinctions among several nuclei, including, for example, the Vim, Ventro-odalis (Vo), and Latero-polaris (Lpo). These findings were consistent across subjects 1, 3 and 4 (*Supp. Fig. 3*, *Supp. Fig. 4*). Subject 2 displayed suboptimal QSM quality, yet the nuclei remained distinguishable (*Supp. Fig. 2*). In parallel, among the observed sequences, GWM$_{opt}$ yielded the sharpest contrast between nuclei such as the Parafascicular (Pf) and Centromedian (Cm) nuclei.

While most nucleus pairs could be differentiated using at least one image contrast, there remained some pairs that eluded distinction across all subjects. According to the Morel-based atlas, none of the contrasts proved effective in distinguishing between AV and MD-Pf, VA and VLP$_{sup}$, as well as VLP$_{sup}$ and VPL. In the case of the Schaltenbrand atlas, distinguishing between the pairs Cm and Ventro-odalis internus (Voi), Lamella medialis (Lam) and Pf, and Lateral and Medial Pulvinar (Pul, Pum), proved to be unattainable.

[see Table 3 on page 30]

## 3.3 Quantitative analysis

Upon fitting our sigmoid-gaussian model to the nuclei pair profiles, more quantitative insights could be obtained to support the qualitative analysis. In specific cases, the profile analysis effectively captured intensity variations across nuclei in the form of a step function, such as for instance between the Ventro-caudalus (Vc) and Pum, between the Pum and Pul, and between the Cm and Pf, with GWM$_{opt}$ (*Fig. 4*). A strong step behavior was also observed in QSM (both QSM$_{COS}$ and QSM$_{STAR}$) between the Vo and Vim and between the Pf and Cm nuclei, for instance, which agreed with the qualitative analysis. Other pairs of nuclei which, particularly on the susceptibility sequences, possessed a darker contour, did accordingly show a small intensity dip, which was suitably modeled by the gaussian function, as exemplified between the Vc and Pum/Pul, and between the Lpo and Vo, for QSM$_{COS}$. In general, this darker contour was not equally well observed in QSM$_{STAR}$, possibly due to the increased smoothness promoted by the regularization terms included in STAR-QSM. Conversely, in cases of low visual distinguishability between nuclei, as seen for instance in T$_1$-based sequences between the Pf and Lam, and between the Pul and Vc, the profile analysis did accordingly result in poorer fits.

[see *Fig. 4* on page 25]

Across contrasts, a clear overall tendency for higher R$^2$ values in GWM$_{opt}$ and QSM was observed for most nuclei pairs, in all subjects (*Fig. 5*, *Supp. Fig. 5*, *Supp. Fig. 6*, *Supp. Fig. 7*, *Supp. Fig. 8*). These results were consistent with our qualitative analysis, and held true for subject 2 as well, despite their lower-quality QSM data. GWM$_{opt}$ outperformed the other T$_1$-based sequences, and excelled in distinguishing the Pul and Pum from other nuclei, and differentiating the Cm from Pf nuclei, for instance. In general, the profiles showed a similar trend across the $T_1$-based sequences, reflecting the general medial-to-lateral differentiation ability already observed qualitatively. Regarding QSM, and in agreement with the qualitative analysis, it was observed that $QSM_{COS}$, on average, exhibited higher $R^2$ values than $QSM_{STAR}$, in all subjects except for subject 2 (*Fig. 5*, *Supp. Fig. 5*, *Supp. Fig. 6*, *Supp. Fig. 7*, *Supp. Fig. 8*). Both QSM sequences consistently outperformed SWI. In general, QSM $R^2$ values were substantially higher across various pairs of nuclei, with subject 1 showing the highest values, consistent with their superior image quality among the QSM images. Conversely, subject 2's QSM $R^2$ values were notably lower. Other sequences ($R_2$*map, SWI, $T_2$w and IRTSE) did show generally lower $R^2$ values. Certain deviations from our qualitative analysis included the notably high $R^2$ values observed for the Pum and Pul nuclei. This phenomenon could be linked to the presence of an intensity gradient which the model could fit effectively, although a distinct boundary between the two nuclei was not readily discernible – a finding that aligns with existing literature (Mai and Forutan 2012). The $R_2$* map also showed a superior performance to what was observed in the qualitative analysis, achieving the third best mean $R^2$ of the tested sequences. However, it was observed that this map had particularly good contrast in the slice where the quantitative analysis was focused, in comparison to other slices.

Across regions, most imaging contrasts excelled at distinguishing the Mammillothalamic tract (MTT). Here, all imaging modalities were expected to exhibit a profile dominated by a sigmoid-shaped behavior in MTT differentiation, as no image contour or differences in contrast inhomogeneities within the tract were observed. However, in some modalities (such as QSM), this behavior did not manifest, most likely due to the small size of the MTT making it highly sensitive to registration misalignments, shifting the nucleus away from the atlas contour and thus the endpoint of the profile.

# 4. Discussion

In this work, we have conducted an extensive and systematic evaluation of thalamic structural imaging techniques for humans at 7T – all acquired from the same individual brains, to allow more direct comparisons across modalities without anatomical variability confounds. Combining qualitative and quantitative analysis techniques, the study identified the most effective imaging approaches for depicting different thalamic nuclei, providing guidance on the best sequence choices depending on the target(s) of interest. To our knowledge, this is the most comprehensive study to date devoted to in-vivo structural imaging of the thalamus.

## 4.1 Thalamic nuclei imaging modalities

A primary aim of this study was to identify which sequences are most useful for thalamic imaging in general, as well as which ones can best depict specific thalamic nuclei for more targeted applications. In general, most of the sequences implemented in this work, based on previous reports, were found to be reproducible, i.e. to show contrast and anatomical features that were consistent with the respective literature. Moreover, the images were generally consistent across the four studied subjects. Between the sequences, both qualitative and quantitative analyses led to congruent conclusions, highlighting QSM and $GWM_{opt}$ as the most informative techniques, and finding that most of the nuclei can be distinguished with at least one of these two approaches.

Regarding QSM, while $QSM_{COS}$ is considered the gold standard (Deistung, Schweser, and Reichenbach 2017) and the most valuable to study the properties of magnetic susceptibility as a biomarker, the inclusion of the single-orientation $QSM_{STAR}$ variant in this analysis was motivated by its higher suitability for routine clinical applications, considering its shorter scan time (single scan of 9.5 minutes, as opposed to 5–9 orientations × 9.5 minutes in $QSM_{COS}$), and the difficulty for volunteers/patients in sustaining different head rotations. High-resolution susceptibility imaging can be sensitive to motion disturbances, and a long scan, combined with uncomfortable head orientations, can turn out to be counterproductive, which may in fact have been the case for subject 2. In general, QSM using a single orientation is expected to produce less accurate maps than $QSM_{COS}$, but this study nonetheless confirms its ability to provide informative thalamic images for distinguishing most nuclei (*Fig. 6*a). It is important to acknowledge that, in terms of general image quality, the QSM data showed stronger variability across subjects than, e.g., $GWM_{opt}$, and was particularly suboptimal for one of the subjects (subject 2), including both $QSM_{COS}$ and $QSM_{STAR}$. This larger variability likely results from motion and phase perturbations related to respiration, which susceptibility sequences can be particularly sensitive to (Raj et al. 2000; Wen, Cross, and Yablonskiy 2015). This sort of variability motivates the integration of methods for motion and breathing artifact compensation for this type of acquisitions, in order to maximize their robustness across subjects (Jorge et al. 2020). Regarding $GWM_{opt}$, the results were found to be more consistent across subjects, showing only small SNR changes (*Fig. 6*a). The combined acquisition time for the two most informative techniques ($QSM_{STAR}$ and $GWM_{opt}$) at 0.6 mm resolution was approximately 22.5 minutes, making it relatively feasible for clinical applications.

## 4.2 Thalamic microstructure

The thalamus is a complex structure with a mixed composition of GM and WM (Mai and Forutan 2012). Myelin-based histological atlases, such as Schaltenbrand (Schaltenbrand and Wahren 1977), elucidate how the WM fibers distribute within the thalamus. Upon comparing Schaltenbrand's histological images with our $T_1$-weighted contrasts, and especially $GWM_{opt}$ which is optimized for GM-to-WM contrast (Fig. 6b), we observe a similar medio-lateral gradient, which has also previously been described in ex-vivo samples (Lemaire et al. 2019). It is known that the longitudinal relaxation, described by the $T_1$ constant, is mainly influenced by the amount of bound water in the tissue, which is higher in myelinated than unmyelinated regions (Harkins et al. 2016). $T_1$-weighted sequences may therefore predominantly reflect the amount of myelin within the thalamus. The presence of myelinated fibers is higher in more lateral nuclei, such as the Vc, the MTT and the Habenular (Hb) (Jones 1985; Mai and Forutan 2012). Accordingly, in the case of the $WM_{null}$ approach, the sequence inversion time was set at the expected nulling point for WM, and indeed this strongly suppressed the more lateral thalamic regions, compared to the medial (*Fig. 2*). In $GWM_{opt}$, the two inversion times from the MP2RAGE sequence are tuned to maximize the contrast between GM and WM in the combined image (Marques and Gruetter 2013), which in this framework results instead in hypointense GM relative to WM, but essentially the same medial-to-lateral differentiation (*Fig. 2*).

On the other hand, QSM displayed rich, unique features in the thalamus that were not captured by the $T_1$-based modalities like $GWM_{opt}$. Contrast derived from susceptibility-based images in living organisms can notably be affected by iron content, given that certain naturally occurring iron-containing molecules are paramagnetic (Beard, Connor, and Jones 2009). Indeed, techniques such as SWI and QSM have been experimentally found to closely match histological iron-based stainings (Sun et al.

2015; Shin et al. 2021). Although there is limited publicly available histological data on iron distribution in the thalamus, existing images (Naidich 2013) also show a remarkable spatial agreement with the features found here in QSM and SWI, highlighting structures like the Pu, Pf and Vo nuclei (Fig. *6*b). Moreover, the thalamus is particularly rich in non-heme iron, with changes across age and disease (Treit et al. 2021). Altogether, this evidence strongly suggests that iron distribution in the thalamus may be a primary source for this contrast.

Nonetheless, susceptibility-related contrast is also influenced by other sources, such as myelin and certain aspects of microstructure (Hametner et al. 2018). Myelin is diamagnetic and hence, contrary to iron, produces a decrease in intensity in QSM images. This was the case, for example, for the MTT, which is indeed composed of myelin fiber bundles. Being a major relay structure of the brain with afferent and efferent connectivity to other areas, the thalamus features a system of myelinated fibers that divide its different subparts. Because myelin sheaths have a highly regular cylindrical structure, poor in water, the QSM contrast is sensitive to fiber orientation with respect to $B_0$ (Birkl et al. 2021). The Lamina medullaris interna has been widely studied (Mai and Forutan 2012) and divides the thalamus into the medial, lateral and anterior groups. While there are other secondary WM tracts that divide and encapsulate certain nuclei, the literature describing these tracts is limited (Jones 1985). In line with the initial hypothesis proposed by (Deistung et al. 2013), the hypointense borders separating nuclei such as the Vc and Pu, which we can observe in our QSM images (*Fig. 3*), may correspond to these fibers and affected by their different orientations.

The $R_2^*$ maps, although being related to susceptibility as well, did not exhibit strong contrast features within the thalamus. Interestingly, both QSM and $R_2^*$ are expected to exhibit increased signal intensity in the presence of iron, but when myelin is present, $R_2^*$ increases while QSM decreases (Lee et al. 2012; Langkammer et al. 2010; Schweser et al. 2011; C. Liu et al. 2011; Treit et al. 2021; Hametner et al. 2018; Deistung et al. 2013). It could be speculated that the diminished contrast features in $R_2^*$ might stem from the concurrent impact of both iron and myelin "pushing" the signal in the same direction.

As for $T_2$w images, they are not expected to be substantially sensitive to iron or myelin in comparison to other techniques (Stüber, Pitt, and Wang 2016; Heath et al. 2018). The IRTSE sequence, however, was proposed for the thalamus by (Kanowski et al. 2014), showing good differentiation performance especially in the anterior nuclei — a region of particular interest due to its challenging visualization with other sequences. Unfortunately, in our implementation, these results could not be replicated. It is possible that our approach based on the 3D SPACE sequence may not be capturing the determinant contrast mechanisms of the implementation proposed by the authors, but this remains inconclusive as no mechanistic explanation for this contrast is available.

## 4.3 Challenges and limitations

An important limitation of this study was the absence of an individual ground truth for the nuclei anatomy. An ideal approach to achieve such a reference would entail acquiring MR images in-vivo and, later, performing histological analyses on the same brains post-mortem, which is nearly impossible to carry out in healthy humans. Moreover, the comparison between in-vivo images and post-mortem preparation would still be prone to confounds from associated morphological changes, imaging-related distortions, and registration errors. Thus, a comparison with standard atlases was pursued instead.

Our image analysis, including a qualitative and a quantitative approach, was designed to tackle the challenges described in detail in Methods. In the existing literature, a few examples have attempted quantitative measures for intra-thalamic assessment. However, these efforts are typically constrained to either specific sequences (Tourdias et al. 2014) or involve the analysis of a few sequences with a focus on a single nucleus (J. Li et al. 2020) – and as far as we are aware, mostly focusing on differences between nuclei intensity, and not the presence of discernible borders.

One limitation of the quantitative analysis was that it was conducted on a single slice, instead of the entire thalamus. Our objective was to keep the analysis (namely the profile extraction and model fits) easier to assess visually, for quality control – and even then, the behavior of the sigmoid-gaussian model was not always straightforward to interpret, possibly due to excessive flexibility. In Fig. *1*c, we present three profiles corresponding to ideal scenarios. However, when analyzing real profiles affected by noise and with limited data samples (spatial resolution), the high number of degrees of freedom in the model, even with defined boundary conditions, can potentially lead to overfitting. An additional challenge to this model was the fact that the intensity within certain nuclei is not fully homogenous; for instance, within the Pu, the intensity in $T_1$-weighted sequences was characterized by a gradient without a clearly defined transition between subnuclei, which is consistent with myelin-staining histological data that demonstrates similar behavior of myelin in this nucleus (Schaltenbrand and Wahren 1977). Nonetheless, despite these challenges, our extensive review of the extracted 2D profiles and model fits indicated that, at least for this slice and image contrasts, the model exhibited a favorable trade-off between flexibility and overfitting risk. It is also important to note that the quantitative method is not intended to replace the qualitative approach; rather, the two complement each other for the overall analysis and interpretation.

Another important point of discussion is the practical impact of adding both QSM and GMWM$_{opt}$ to a scanning protocol. While the combined acquisition time of 22.5 min may be well accommodated in a study (or clinical exam) dedicated to the thalamus, it may prove cumbersome when added to an already lengthy fMRI session for the study of e.g. cortical and subcortical connectivity. In the latter case, however, it may be possible to save time by deriving standard $T_1$-weighted anatomical information from the GMWM$_{opt}$ acquisition, avoiding the need for a dedicated standard $T_1$ scan. On its own, the GMWM$_{opt}$ contrast is considerably different from the standard $T_1$w contrast, and tends to yield poor outcomes with tools such as Freesurfer (Fischl 2012). However, an $R_1$ map can be generated from GMWM$_{opt}$ and a (much quicker) $B_1$ map using dictionary matching approaches (see code available in https://github.com/JosePMarques/MP2RAGE-related-scripts, and snippet in Supplementary Material). In turn, this $R_1$ map, based on our experience, appears to be able to replace a standard $T_1$w as input to segmentation tools such as Freesurfer and SynthSeg (Billot et al. 2023).

## 4.4 Future perspectives

Stemming from this study, new lines of research could be interesting to explore. For instance, to compare to the structural information, it would be interesting to obtain fMRI data from the same subjects, to provide insights from functional connectivity, which can also be used to delineate thalamic clusters (Hale et al. 2015; Chen et al. 2021).

On the methodological side, our results highlight the value of QSM and suggest it could be useful to invest further in this modality as input for thalamic segmentation. State-of-the-art methods (Iglesias et al. 2018; Su et al. 2019) have instead relied on $T_1$-based contrasts, which exhibit a distinct medial-to-lateral variation within the thalamus. QSM, however, also allows distinguishing unique contrast features in the anterior-to-posterior direction, especially more laterally, critical for nuclei such as the Vim. Thus, it appears highly relevant to test and compare the more established $T_1$-based methods with more recent emerging approaches such as based on the MuSuS atlas, for example, which relies on QSM information (He et al. 2022). Specific toolboxes for susceptibility mapping contrasts, such as SEPIA (Chan and Marques 2021), could be useful to perform matching-contrast algorithms and atlas co-registrations of this specific atlas to individual subjects.

Another interesting possibility stems from the recognition that $GWM_{opt}$ and QSM both proved to be very informative, delivering non-redundant contrast features. Exploring the combination/fusion of these images into one could potentially enhance the visualization of thalamic nuclei and simplify their identification. The $GWM_{opt}$ contrast may primarily arise from myelin content, whereas the QSM contrast may be closely linked to iron. Combining $T_1w$ and QSM values into a single output has been previously explored in simpler ex-vivo samples, including a straightforward linear combination of both sequences (Dadarwal, Ortiz-Rios, and Boretius 2022). Analogous efforts could be explored in-vivo for $GWM_{opt}$ and QSM, with atlas information as a reference, for example.

In another perspective, instead of combining different acquisitions into a single image, the multi-contrast information could instead be jointly used as input for automatic segmentation. Current methods for automatic thalamic segmentation are based on a single image source (He et al. 2022; Vidal et al. 2024; Su et al. 2019; Iglesias et al. 2018) and have shown some limitations in accuracy (Williams, Roesch, and Christakou 2022), suggesting room for important improvements. New multi-contrast segmentation methods could range from standard techniques like k-means clustering to more advanced methodologies. One notable example in the literature used $T_1$-weighted, $T_2$*-weighted and QSM contrasts for supervised classification with a multi-class convex segmentation technique, to produce a relatively coarse differentiation between three nuclei groups (lateral, medial and posterior) (Corona et al. 2020). Based on the differentiation ability we have observed with the sub-millimeter sequences explored in this work, it appears likely that substantially finer segmentations could be developed with this strategy.

## 5. Conclusion

To the best of our knowledge, this study represents the most comprehensive assessment of structural MRI contrasts for thalamic imaging to date. The data revealed that QSM and $T_1$-weighted MP2RAGE optimized for GM-to-WM contrast are currently the most valuable options, enabling the delineation of numerous thalamic nuclei across all participants. These findings also point to promising avenues of improvement for imaging-based thalamic nuclei segmentation.

# Acknowledgments

This work was funded by the Swiss National Science Foundation through grant PZ00P2_185909, and supported by CSEM – Swiss Center for Electronics and Microtechnology, by the Swiss Institute for Translational and Entrepreneurial Medicine (SITEM), and by the CIBM Center for Biomedical Imaging, Switzerland.

# Disclosure/conflicts of interest

Tom Hilbert and Gabriele Bonanno are employed by Siemens Healthineers International AG, Switzerland.

# Data availability statement

The comprehensive multi-contrast imaging data collected in this study will be made publicly available upon acceptance of this manuscript as a peer-reviewed publication.

# Figures

## a) Registration

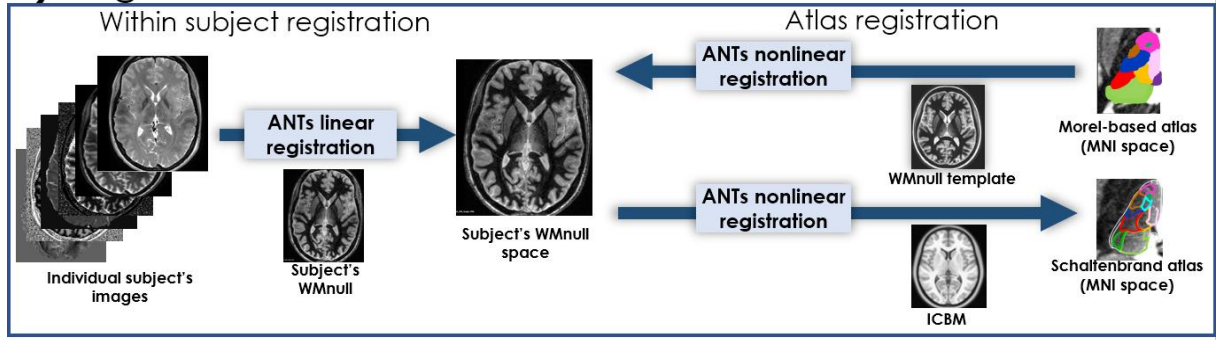

## b) Qualitative analysis

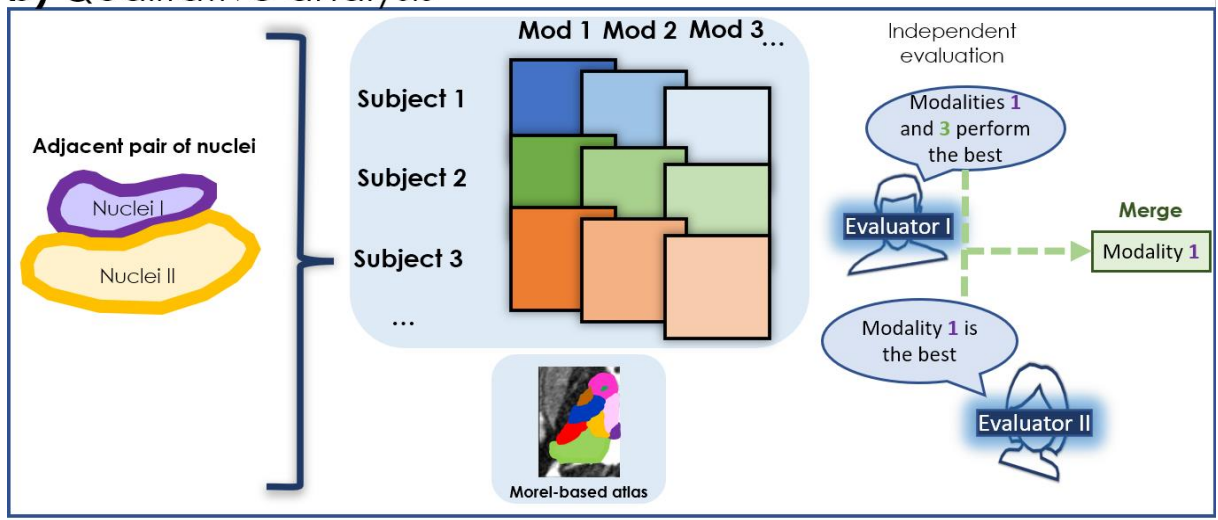

## c) Quantitative analysis

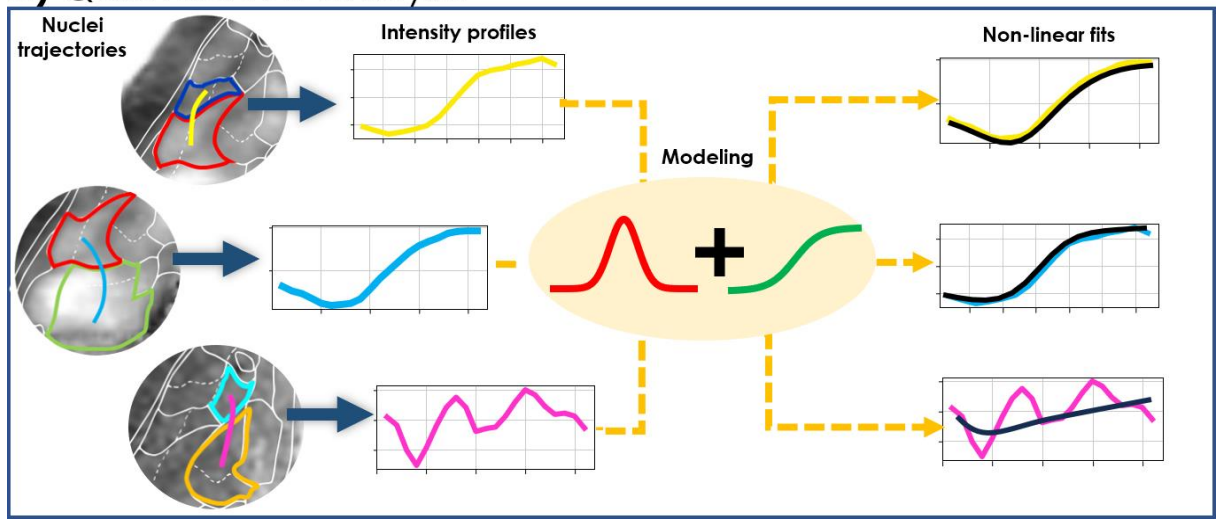

Fig. 1. Schematic outline of data processing and analysis. a) Registration, encompassing within-subject and atlas registration to the $WM_{null}$ space. b) Qualitative analysis procedure, involving independent evaluations followed by a consensus merging. c) Quantitative analysis, illustrated for three thalamic pair examples with distinct intensity profiles across the nuclei; the first two cases show good distinguishably, which translates into good fits of the gaussian-sigmoid model.

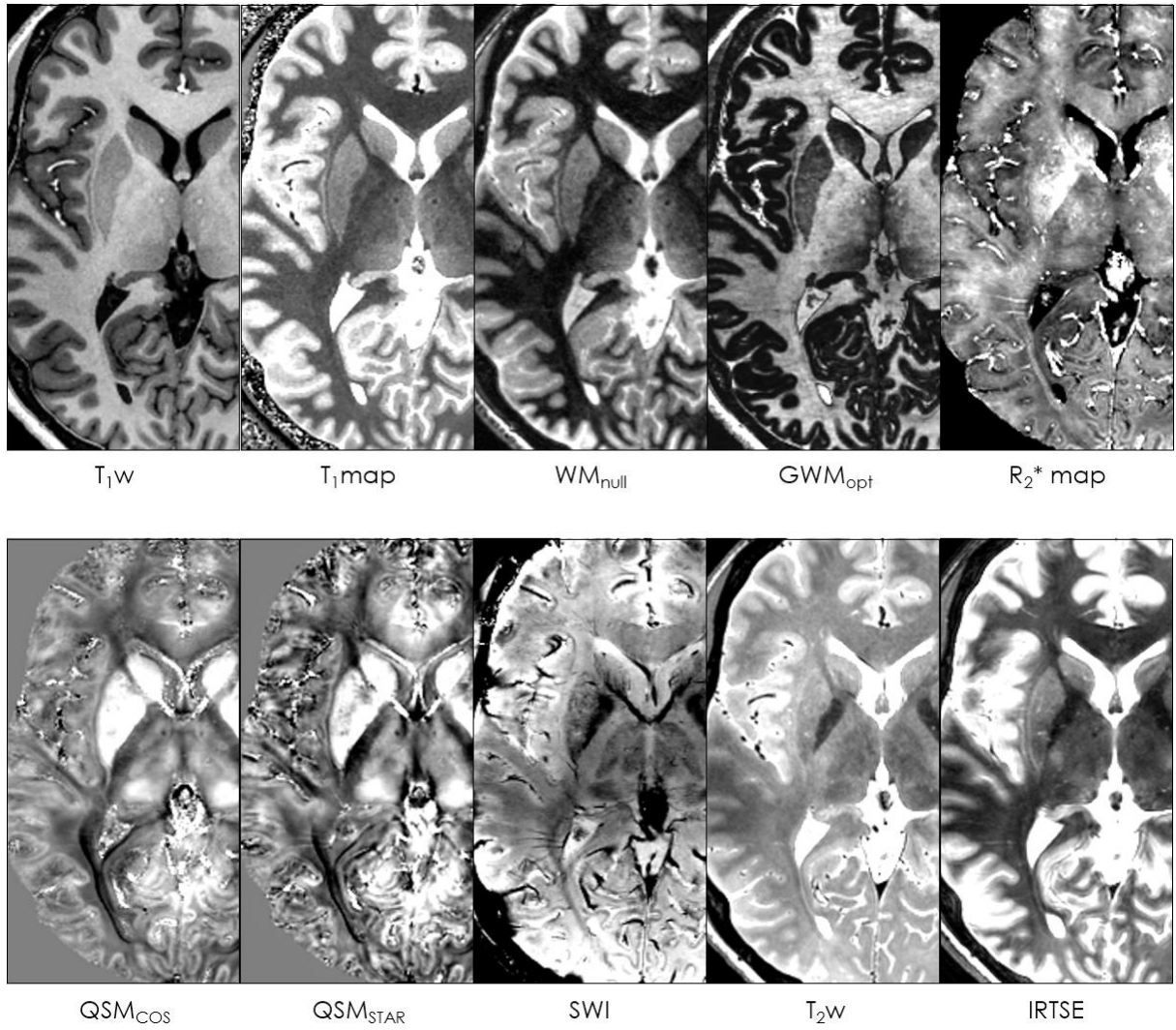

Fig. 2. Overview of the diverse MRI contrasts acquired in this study, for subject 1.

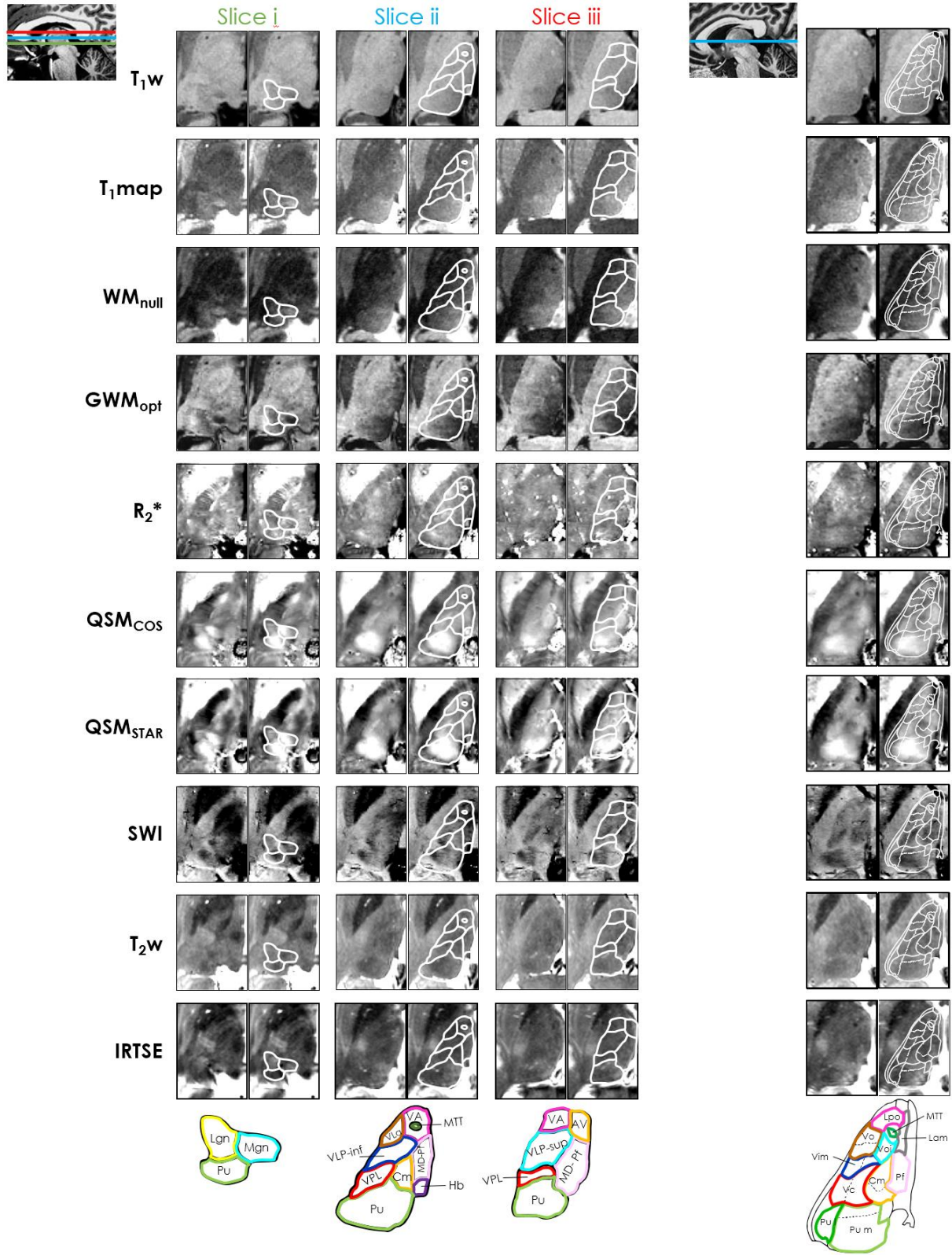

Fig. 3. Thalamic in-vivo images (representative axial slices) compared against Morel-based (left) and Schaltenbrand atlases (right), for subject 1. The anatomical location of each slice is displayed on the upper left corner. For the Schaltenbrand atlas in MNI space, the selected slice is z = 3.5mm. Each in-vivo image is shown with and without an atlas overlay. The labels are provided at the bottom. Legend (left): AV: Anterior Ventral, VA: Ventral Anterior, VLa: Ventral Lateral anterior, VLP-inf: Ventral Lateral Posterior-inferior, VPL: Ventral Posterior Lateral, VLP-sup: Ventral Lateral Posterior-superior,

Pu: Pulvinar, Lgn: Lateral geniculate, Mgn: Medial geniculate, Cm: Centromedian, MD-Pf: Mediodorsal-Parafascicular, Hb: Habenular, Mtt: Mammillothalamic tract. (right) Lpo: Latero-polaris, Vo: Ventro-odalis, Voi: Ventro-odalis internus, Vim: Ventral intermediate, Vc: Ventro-caudalus, Pum: Medial pulvinar, Pul: Lateral pulvinar, Cm: Centromedian, Pf: Parafascicular, Lam: Lamella medialis, Mtt: Mammillothalamic tract.

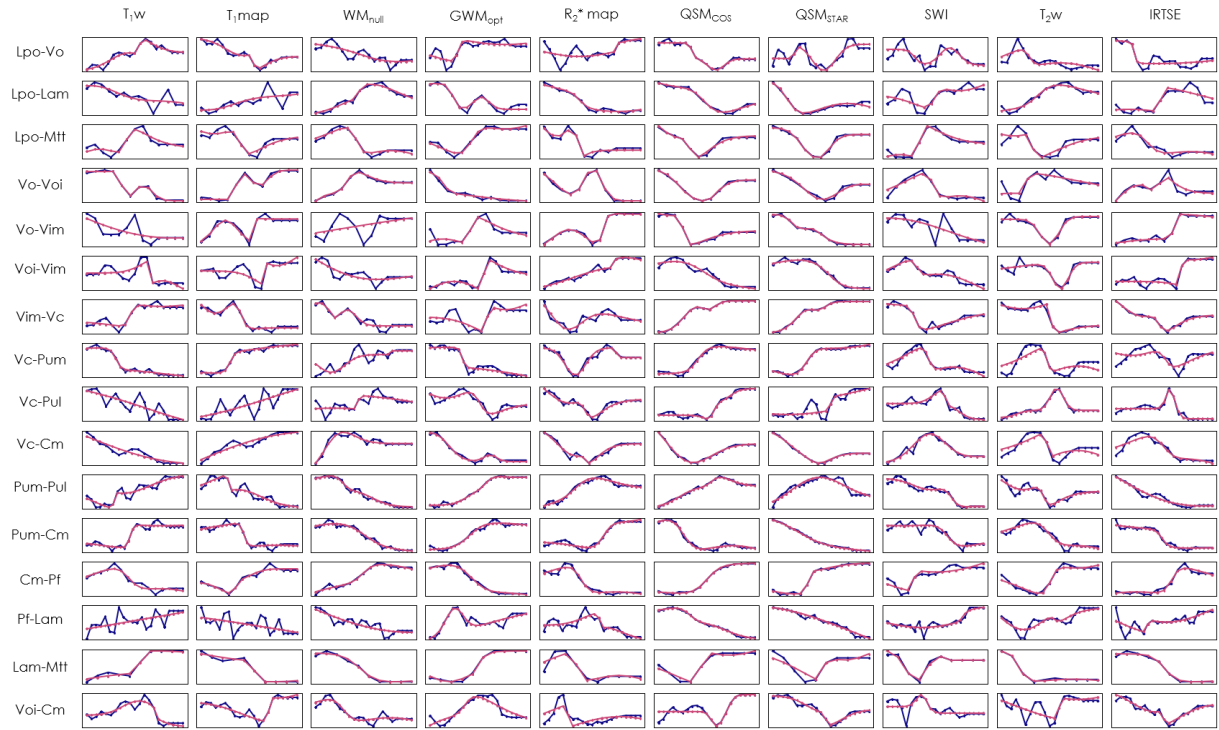

Fig. 4. Intensity profiles of nuclei pairs for subject 1, with reference to the Schaltenbrand atlas at slice z = 3.5mm, for the different MRI contrasts. Each graph shows the intensity profile (blue), and the corresponding model fit (pink).

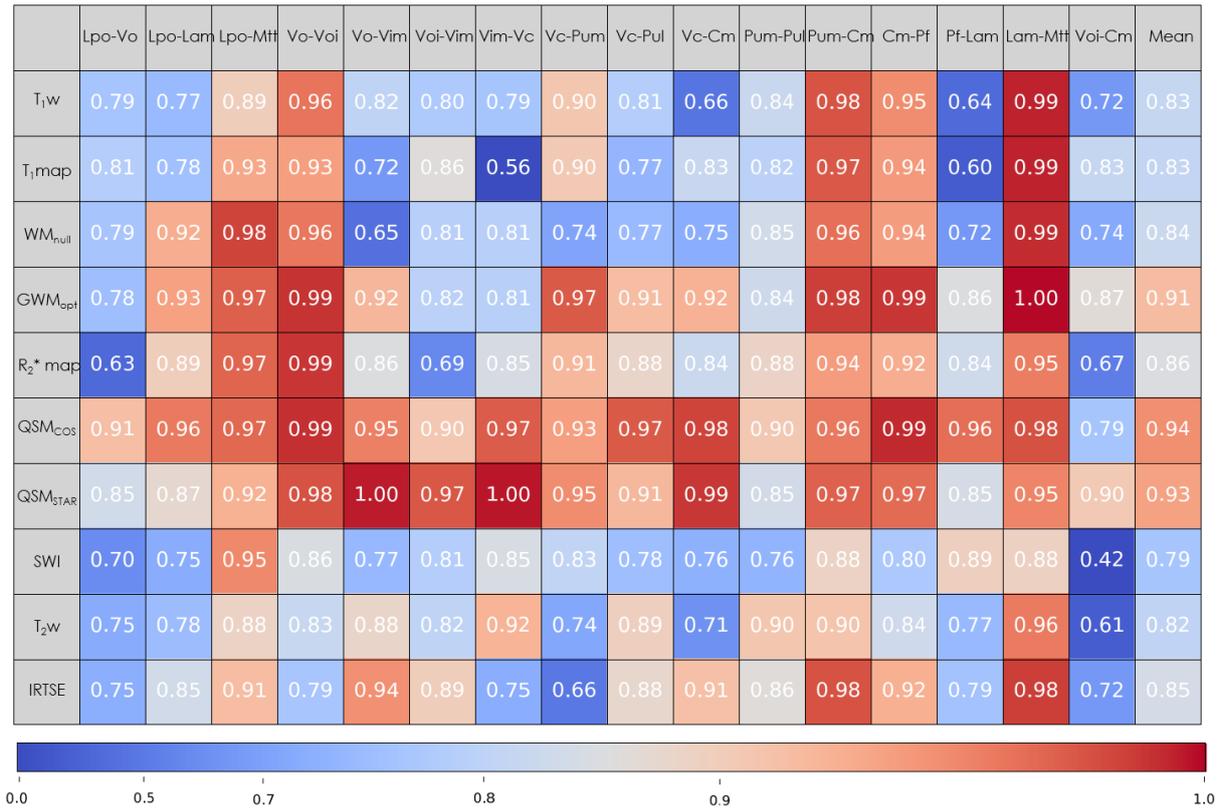

Fig. 5. $R^2$ values obtained from the profile model fits of each pair of nuclei and contrast, averaged across all subjects. The mean $R^2$ value across all nuclei pairs for each contrast is displayed on the right-most column. A matching color (in logarithmic scale) was added behind each corresponding value, to aid visualization.

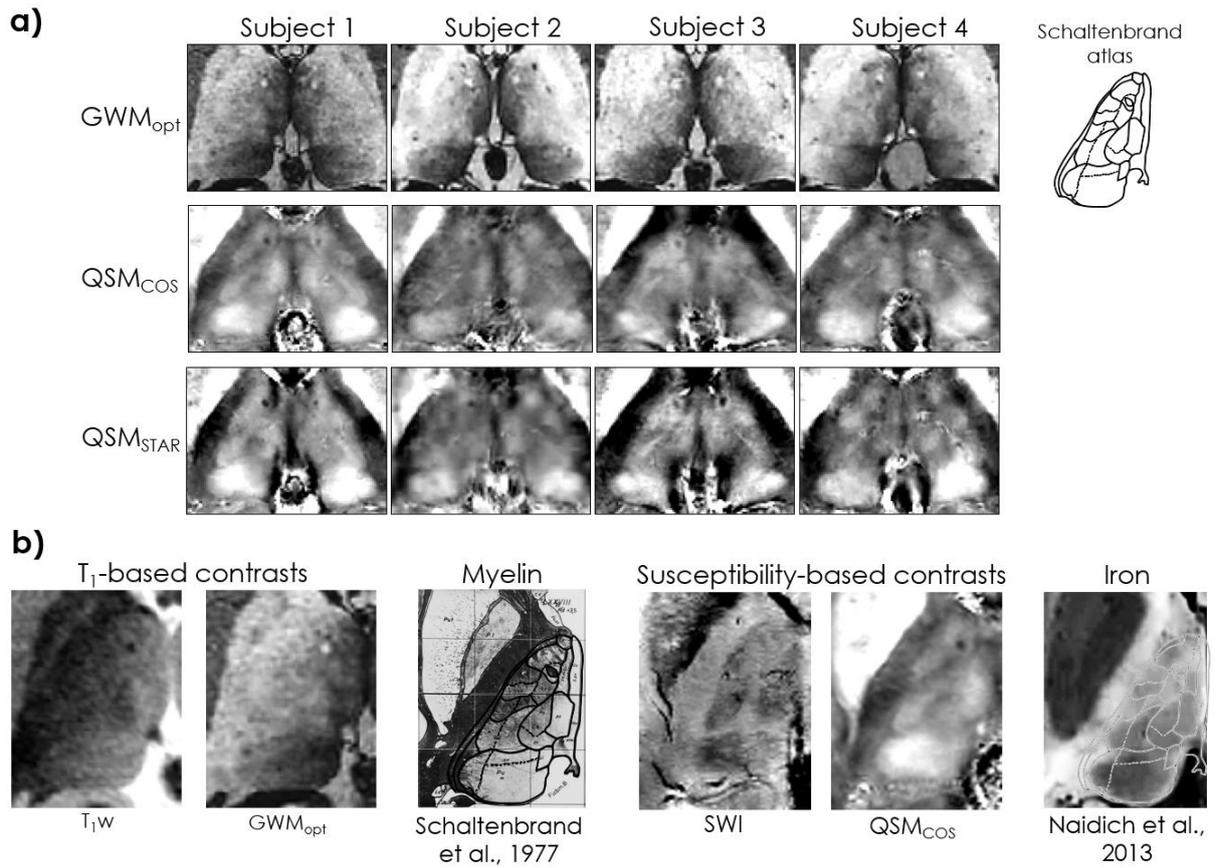

Fig. 6. a) Thalamic contrast for the two most valuable sequences found in this work: GWM$_{opt}$ and QSM (COSMOS and STAR-QSM), for all 4 subjects, showing the respective contrast variability across individuals. b) Spatial agreement between (i) the T$_1$-based sequences and myelin-stain histology (Schaltenbrand and Wahren 1977), and (ii) susceptibility-based sequences and iron-stain histology (Drayer et al. 1986; Naidich 2013). Both histological images were adapted with permission from the respective source publications. The selected slice from the MRI contrast and myelin-stain histology map corresponds to z = 3.5 mm in MNI space, which appears to correspond anatomically to the iron-based slice as well. The Schaltenbrand atlas contour was overlaid on the histology maps.

# Tables

Table 1. MRI sequences and parameters utilized in this study

| Sequence names | Timings (ms) | Flip angles (°) | Resolution (mm) | Readout bandwidth (Hz/Px) | Undersampling | Turbo factor | Acquisition time (minutes) | Generated contrast(s) | Reference |
|---|---|---|---|---|---|---|---|---|---|
| $B_1^+$ mapping 3D SA2RAGE research application sequence | $TE/TD_1/TD_2/TR$ = 1.07/60/1800/2400 | $\alpha_1/\alpha_2 = 4/11$ | 2.0 iso | 490 | 2× GRAPPA (first PE) 6/8 partial Fourier (both PEs) | -- | 2.25 | $B_1^+$ map | (Eggenschwiler et al. 2011) |
| CS-accelerated 3D MP2RAGE research application sequence | $TE/TI_1/TI_2/TR$ = 2.06/800/2700/6000 | $\alpha_1/\alpha_2 = 4/5$ | 0.6 iso | 240 | 4.0× CS | 252 | 7 | $T_1w$, $T_1map$ | (Mussard et al. 2020) |
| 3D MPRAGE tuned to suppress WM signal | $TE/TI/TR$ = 2.12/680/6000 | $\alpha = 4$ | 0.6 iso | 240 | 3× GRAPPA (first PE) 6/8 partial Fourier (both PEs) | -- | 10 | $WM_{null}$ | (Tourdias et al. 2014) |
| CS-accelerated 3D MP2RAGE research application tuned to maximize GM-to-WM contrast | $TE/TI_1/TI_2/TR$ = 2.06/700/1600/6000 | $\alpha_1/\alpha_2 = 7/7$ | 0.6 iso | 240 | 4.0× CS | 140 | 13 | $GWM_{opt}$ | (Marques and Gruetter 2013) |
| Flow-compensated 3D multi-echo GRE | 5 echoes $TE_1/\Delta TE \approx 5.38/5.20$ TR = 31 | $\alpha = 11$ | 0.6 iso | 280 | 2× GRAPPA (both PEs), elliptical scanning | -- | 9.5 | QSM | (Wei et al. 2015; T. Liu et al. 2009) |
| Flow-compensated 3D GRE sequence | TE/TR = 20/28 | $\alpha = 11$ | 0.375×0.375×1 | 240 | 2× GRAPPA (first PE) | -- | 11 | SWI | (Haacke et al. 2009) |
| VFA 3D TSE SPACE | TE/TR = 147/3370 | variable | 0.6 iso | 521 | 2×2 CAIPIRINHA | 96 | 12.5 | $T_2w$ | (Mugler 2014) |
| IR-prepared VFA 3D TSE SPACE | TE/TI/TR = 148/500/5630 | variable | 0.6×0.6×2.0 | 521 | 2×2 CAIPIRINHA | 80 | 6.5 | IRTSE | (Kanowski et al. 2014) |

Table 2. Qualitative analysis: identification of the imaging modality/ies that best differentiate each pair of adjacent thalamic nuclei, for a Morel-based atlas

| Morel-based atlas | | |
|---|---|---|
| **Nucleus 1** | **Nucleus 2** | **Consensus sequence(s)** |
| AV | MD-Pf | *None* |
| | VA | **GWM$_{opt}$** |
| Cm | MD-Pf | **GWM$_{opt}$** **QSM** |
| | Pu | **GWM$_{opt}$** **QSM** |
| | VLP$_{inf}$ | **QSM** |
| | VPL | **QSM** |
| Hb | MD-Pf | **GWM$_{opt}$** **QSM** |
| | Pu | **GWM$_{opt}$** **QSM** |
| Lgn | Mgn | **QSM** **GWM$_{opt}$** |
| | Pu | **QSM** |
| MD-Pf | Pu | **QSM** |
| | VA | **QSM** **SWI** |
| | VLP$_{inf}$ | **QSM** **GWM$_{opt}$** |
| | VLP$_{sup}$ | **GWM$_{opt}$** |
| | VPL | **GWM$_{opt}$** |
| Mgn | Pu | **QSM** **GWM$_{opt}$** |
| Mtt | VA | **QSM** **GWM$_{opt}$** **WM$_{null}$** |
| Pu | VLP$_{sup}$ | **GWM$_{opt}$** |
| | VPL | **GWM$_{opt}$** **QSM** |
| VA | VLa | **QSM** **GWM$_{opt}$** |
| | VLP$_{inf}$ | **GWM$_{opt}$** **QSM** |
| | VLP$_{sup}$ | *None* |
| VLa | VLP$_{inf}$ | **QSM** |
| VLP$_{inf}$ | VLP$_{sup}$ | **GWM$_{opt}$** |
| | VPL | **QSM** **SWI** |
| VLP$_{sup}$ | VPL | *None* |

**Note:** This result represents a consensus between the two evaluators, across the four subjects of the group. Cases where one or both evaluators decided that no modality could offer a clear differentiation are indicated with "None". "QSM" includes both QSM$_{COS}$ and QSM$_{STAR}$.

Table 3. Qualitative analysis: Identification of the imaging modality/ies that best differentiate each pair of adjacent thalamic nuclei, for the Schaltenbrand atlas

| Schaltenbrand atlas | | |
|---|---|---|
| Cm | Pf | **GWM$_{opt}$** |
| | Pu m | **GWM$_{opt}$** **QSM** |
| | Vc | **QSM** |
| | Voi | *None* |
| Lam | Lpo | **GWM$_{opt}$** |
| | Mtt | **QSM** **GWM$_{opt}$** |
| | Pf | *None* |
| Lpo | Mtt | **QSM** **GWM$_{opt}$** |
| | Vo | **GWM$_{opt}$** **QSM** |
| Pu l | Pu m | *None* |
| | Vc | **QSM** **GWM$_{opt}$** |
| Pu m | Vt | **GWM$_{opt}$** **QSM** |
| Vc | Vim | **QSM** |
| Vim | Vo | **QSM** |
| | Voi | **QSM** **GWM$_{opt}$** |
| Vo | Voi | **QSM** |

**Note:** This result represents a consensus between the two evaluators, across the four subjects of the group. Cases where one or both evaluators decided that no modality could offer a clear differentiation are indicated with "None". "QSM" includes both QSM$_{COS}$ and QSM$_{STAR}$.

# Supplementary figures

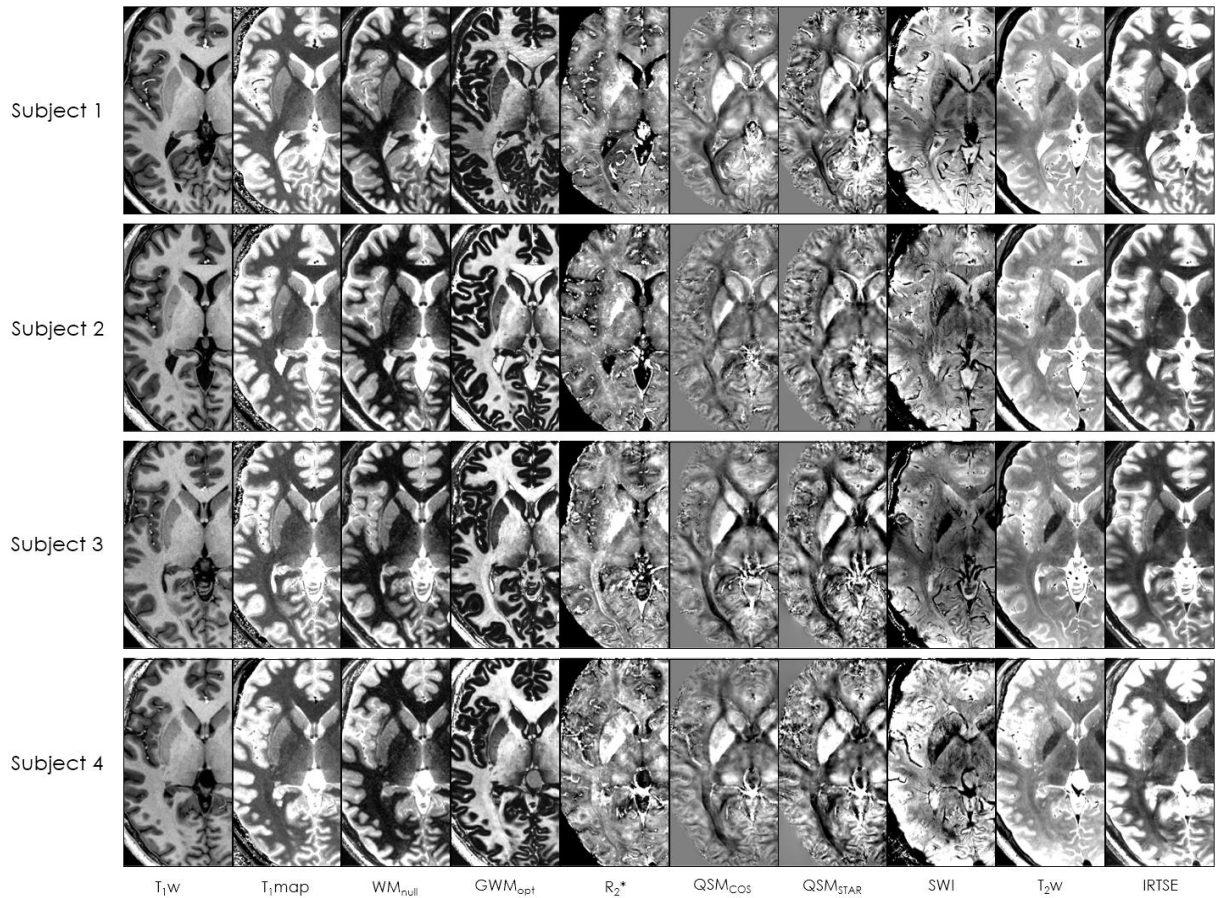

Supp. Fig. 1. Overview of the diverse MRI contrasts acquired in this study, for all subjects.

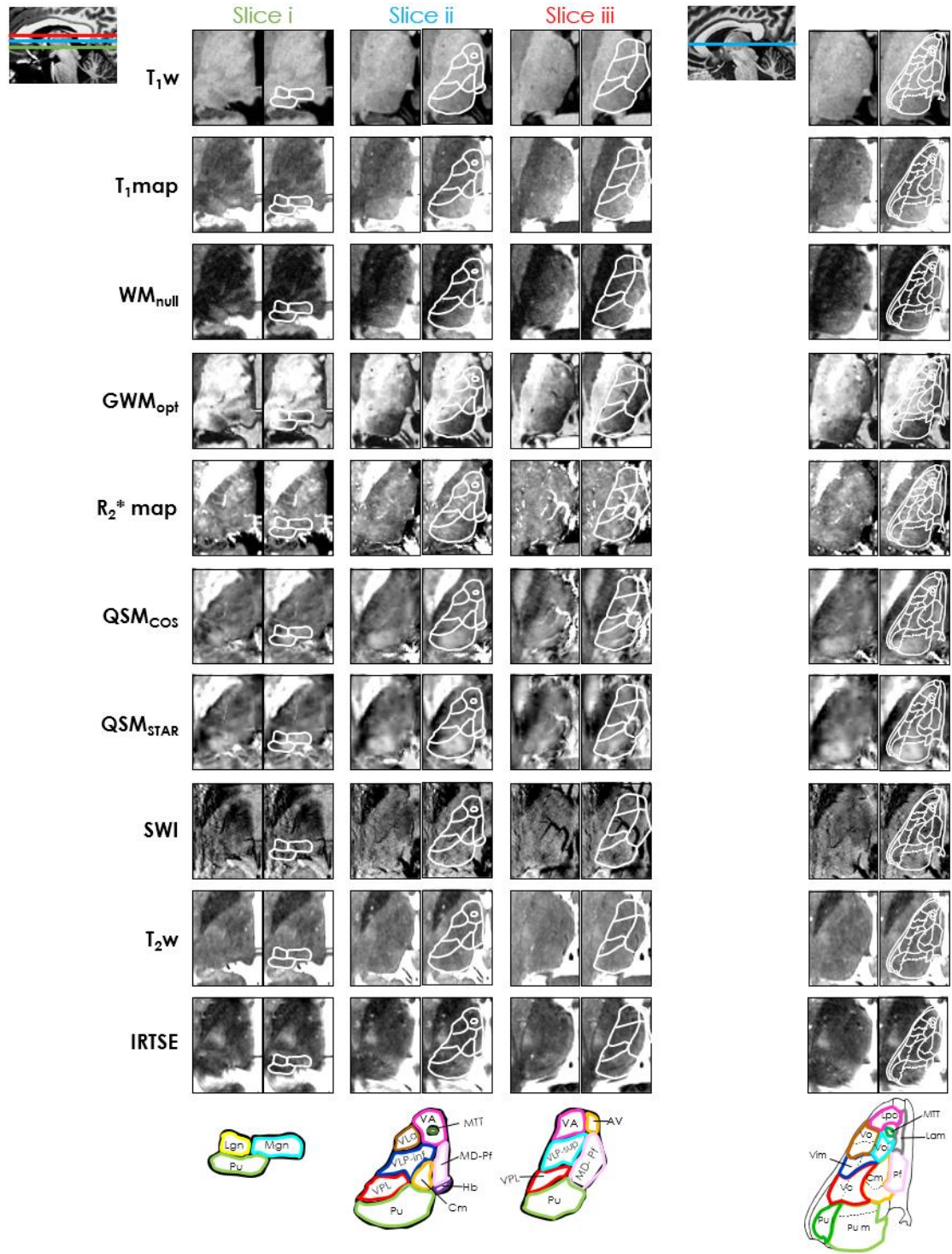

Supp. Fig. 2. Thalamic in-vivo images (representative axial slices) compared against Morel-based (left) and Schaltenbrand atlases (right), for subject 2. The anatomical location of each slice is displayed on the upper left corner. For the Schaltenbrand atlas in MNI space, the selected slice is z = 3.5mm. Each in-vivo image is shown with and without an atlas overlay. The labels are provided at the bottom. Legend (left): AV: Anterior Ventral, VA: Ventral Anterior, VLa: Ventral Lateral anterior, VLP-inf: Ventral Lateral Posterior-inferior, VPL: Ventral Posterior Lateral, VLP-sup: Ventral Lateral Posterior-superior, Pu: Pulvinar, Lgn: Lateral geniculate, Mgn: Medial geniculate, Cm: Centromedian, MD-Pf: Mediodorsal-Parafascicular, Hb: Habenular, Mtt: Mammillothalamic tract. (right) Lpo: Latero-polaris, Vo: Ventro-odalis, Voi: Ventro-odalis internus, Vim: Ventral intermediate, Vc: Ventro-caudalus, Pum: Medial pulvinar, Pul: Lateral pulvinar, Cm: Centromedian, Pf: Parafascicular, Lam: Lamella medialis, Mtt: Mammillothalamic tract.

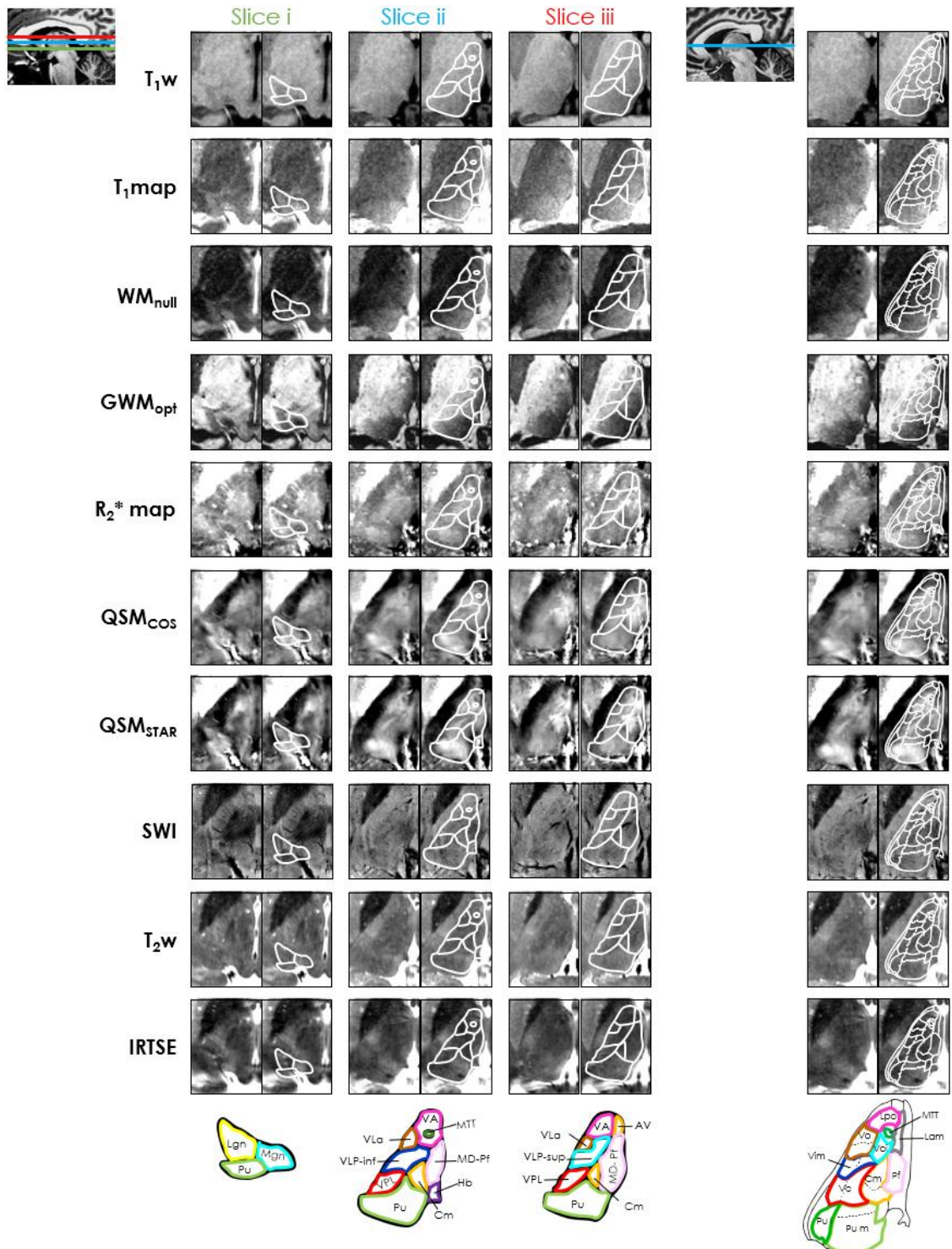

Supp. Fig. 3. Thalamic in-vivo images (representative axial slices) compared against Morel-based (left) and Schaltenbrand atlases (right), for subject 3. The anatomical location of each slice is displayed on the upper left corner. For the Schaltenbrand atlas in MNI space, the selected slice is z = 3.5mm. Each in-vivo image is shown with and without an atlas overlay. The labels are provided at the bottom. Legend (left): AV: Anterior Ventral, VA: Ventral Anterior, VLa: Ventral Lateral anterior, VLP-inf: Ventral Lateral Posterior-inferior, VPL: Ventral Posterior Lateral, VLP-sup: Ventral Lateral Posterior-superior, Pu: Pulvinar, Lgn: Lateral geniculate, Mgn: Medial geniculate, Cm: Centromedian, MD-Pf: Mediodorsal-Parafascicular, Hb: Habenular, Mtt: Mammillothalamic tract. (right) Lpo: Latero-polaris, Vo: Ventro-odalis, Voi: Ventro-odalis internus, Vim: Ventral intermediate, Vc: Ventro-caudalus, Pum: Medial pulvinar, Pul: Lateral pulvinar, Cm: Centromedian, Pf: Parafascicular, Lam: Lamella medialis, Mtt: Mammillothalamic tract.

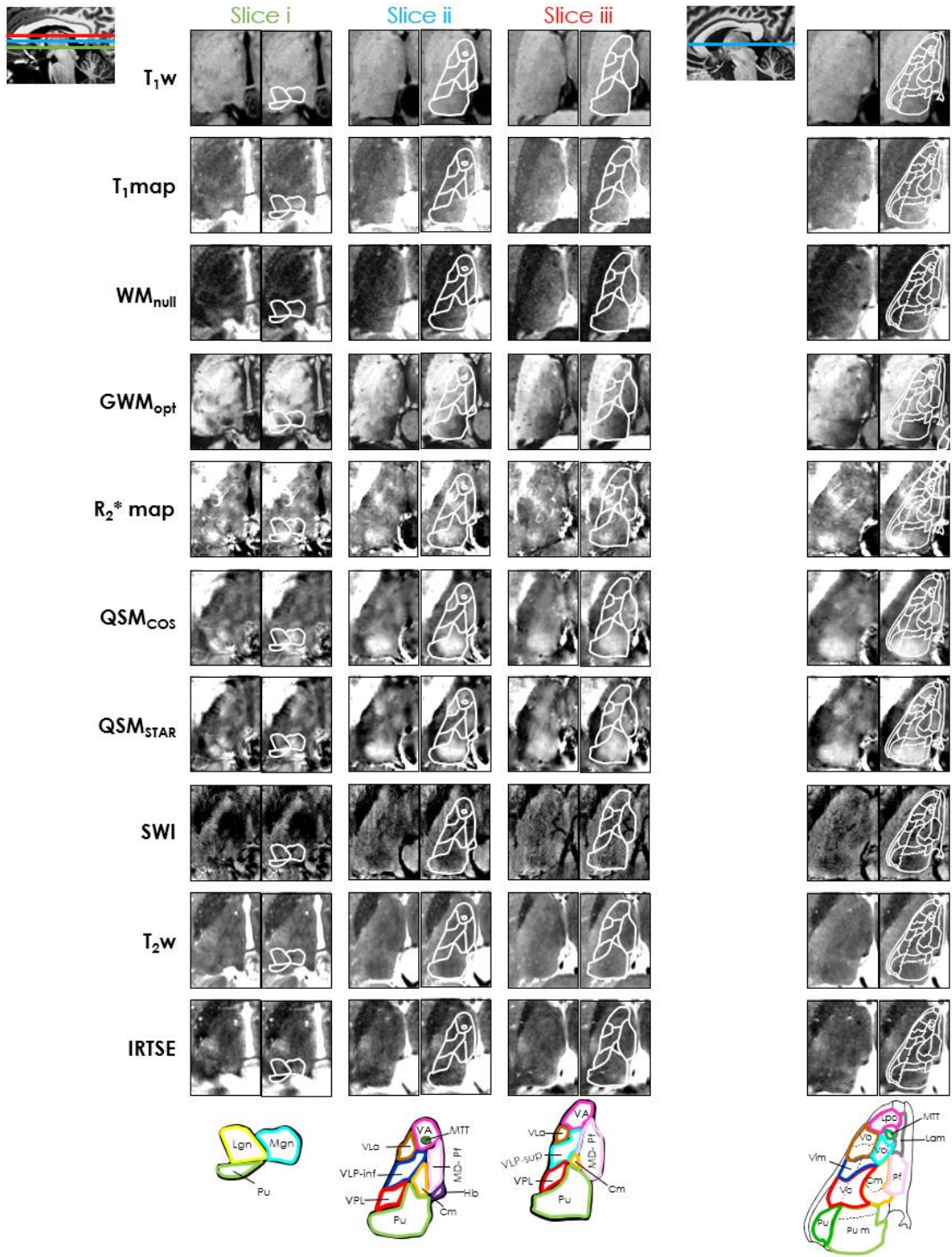

Supp. Fig. 4. Thalamic in-vivo images (representative axial slices) compared against Morel-based (left) and Schaltenbrand atlases (right), for subject 4. The anatomical location of each slice is displayed on the upper left corner. For the Schaltenbrand atlas in MNI space, the selected slice is z = 3.5mm. Each in-vivo image is shown with and without an atlas overlay. The labels are provided at the bottom. Legend (left): AV: Anterior Ventral, VA: Ventral Anterior, VLa: Ventral Lateral anterior, VLP-inf: Ventral Lateral Posterior-inferior, VPL: Ventral Posterior Lateral, VLP-sup: Ventral Lateral Posterior-superior, Pu: Pulvinar, Lgn: Lateral geniculate, Mgn: Medial geniculate, Cm: Centromedian, MD-Pf: Mediodorsal-Parafascicular, Hb: Habenular, Mtt: Mammillothalamic tract. (right) Lpo: Latero-polaris, Vo: Ventro-odalis, Voi: Ventro-odalis internus, Vim: Ventral intermediate, Vc: Ventro-caudalus, Pum: Medial pulvinar, Pul: Lateral pulvinar, Cm: Centromedian, Pf: Parafascicular, Lam: Lamella medialis, Mtt: Mammillothalamic tract.

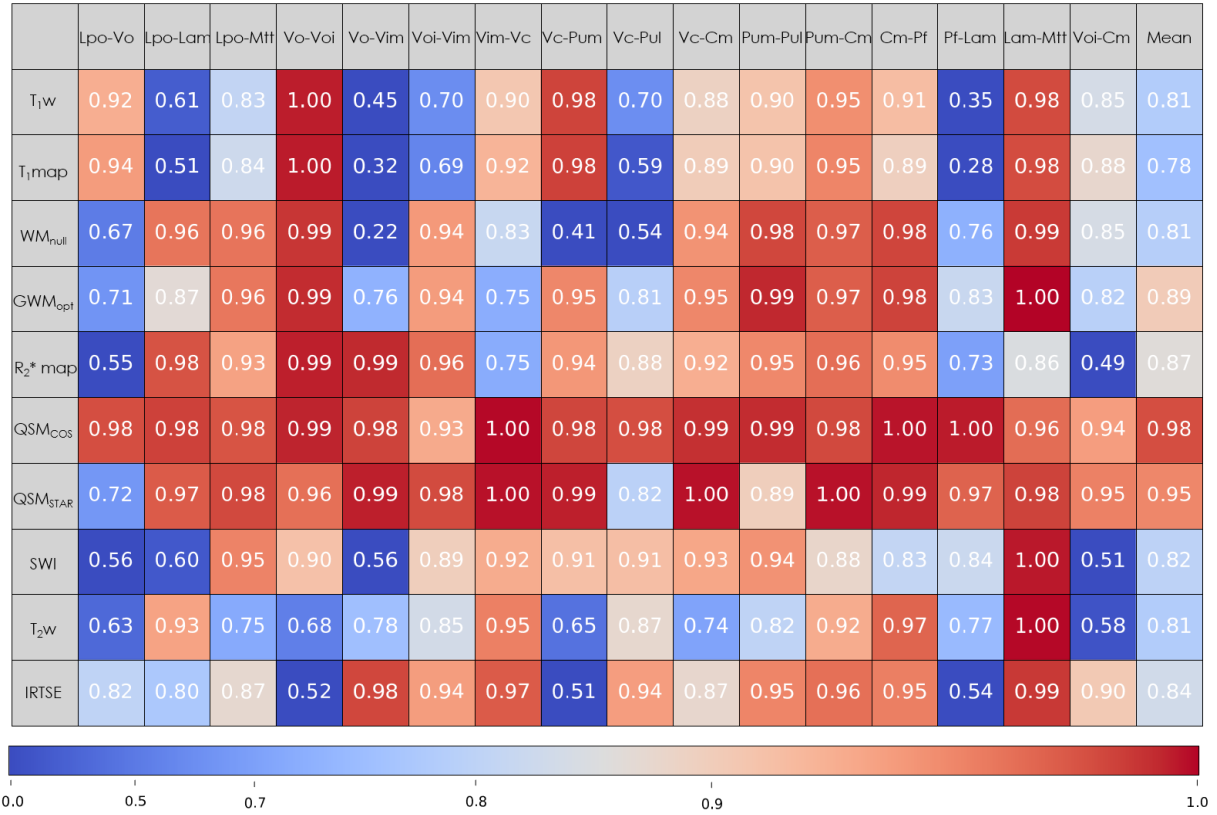

| | Lpo-Vo | Lpo-Lam | Lpo-Mtt | Vo-Voi | Vo-Vim | Voi-Vim | Vim-Vc | Vc-Pum | Vc-Pul | Vc-Cm | Pum-Pul | Pum-Cm | Cm-Pf | Pf-Lam | Lam-Mtt | Voi-Cm | Mean |
|---|---|---|---|---|---|---|---|---|---|---|---|---|---|---|---|---|---|
| T$_1$w | 0.92 | 0.61 | 0.83 | 1.00 | 0.45 | 0.70 | 0.90 | 0.98 | 0.70 | 0.88 | 0.90 | 0.95 | 0.91 | 0.35 | 0.98 | 0.85 | 0.81 |
| T$_1$map | 0.94 | 0.51 | 0.84 | 1.00 | 0.32 | 0.69 | 0.92 | 0.98 | 0.59 | 0.89 | 0.90 | 0.95 | 0.89 | 0.28 | 0.98 | 0.88 | 0.78 |
| WM$_{null}$ | 0.67 | 0.96 | 0.96 | 0.99 | 0.22 | 0.94 | 0.83 | 0.41 | 0.54 | 0.94 | 0.98 | 0.97 | 0.98 | 0.76 | 0.99 | 0.85 | 0.81 |
| GWM$_{opt}$ | 0.71 | 0.87 | 0.96 | 0.99 | 0.76 | 0.94 | 0.75 | 0.95 | 0.81 | 0.95 | 0.99 | 0.97 | 0.98 | 0.83 | 1.00 | 0.82 | 0.89 |
| R$_2$* map | 0.55 | 0.98 | 0.93 | 0.99 | 0.99 | 0.96 | 0.75 | 0.94 | 0.88 | 0.92 | 0.95 | 0.96 | 0.95 | 0.73 | 0.86 | 0.49 | 0.87 |
| QSM$_{Cos}$ | 0.98 | 0.98 | 0.98 | 0.99 | 0.98 | 0.93 | 1.00 | 0.98 | 0.98 | 0.99 | 0.99 | 0.98 | 1.00 | 1.00 | 0.96 | 0.94 | 0.98 |
| QSM$_{STAR}$ | 0.72 | 0.97 | 0.98 | 0.96 | 0.99 | 0.98 | 1.00 | 0.99 | 0.82 | 1.00 | 0.89 | 1.00 | 0.99 | 0.97 | 0.98 | 0.95 | 0.95 |
| SWI | 0.56 | 0.60 | 0.95 | 0.90 | 0.56 | 0.89 | 0.92 | 0.91 | 0.91 | 0.93 | 0.94 | 0.88 | 0.83 | 0.84 | 1.00 | 0.51 | 0.82 |
| T$_2$w | 0.63 | 0.93 | 0.75 | 0.68 | 0.78 | 0.85 | 0.95 | 0.65 | 0.87 | 0.74 | 0.82 | 0.92 | 0.97 | 0.77 | 1.00 | 0.58 | 0.81 |
| IRTSE | 0.82 | 0.80 | 0.87 | 0.52 | 0.98 | 0.94 | 0.97 | 0.51 | 0.94 | 0.87 | 0.95 | 0.96 | 0.95 | 0.54 | 0.99 | 0.90 | 0.84 |

| 0.0 | 0.5 | 0.7 | 0.8 | 0.9 | 1.0 |

Supp. Fig. 5. $R^2$ values obtained from the profile model fits of each pair of nuclei and contrast for subject 1. The mean $R^2$ value across all nuclei pairs for each contrast is displayed on the right-most column. A matching color (in logarithmic scale) was added behind each corresponding value, to aid visualization.

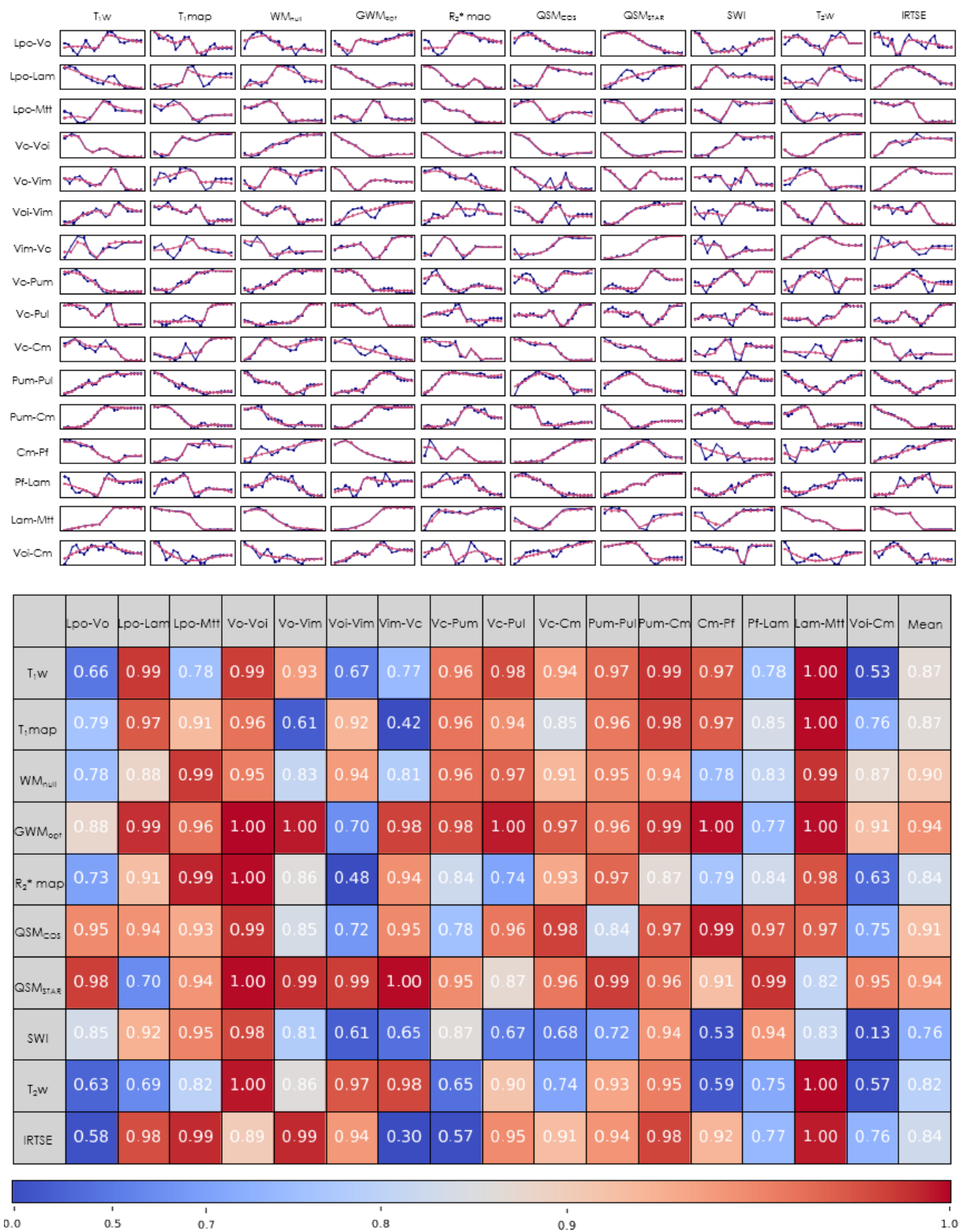

Supp. Fig. 6. Top: Intensity profiles of nuclei pairs for subject 2, with reference to the Schaltenbrand atlas at slice z = 3.5mm, for the different MRI contrasts. Each graph shows the intensity profile (blue), and the corresponding model fit (pink). Bottom: $R^2$ values obtained from the profile model fits of each pair of nuclei and contrast for subject 2. The mean $R^2$ value across all nuclei pairs for each contrast is displayed on the right-most column. A matching color (in logarithmic scale) was added behind each corresponding value, to aid visualization.

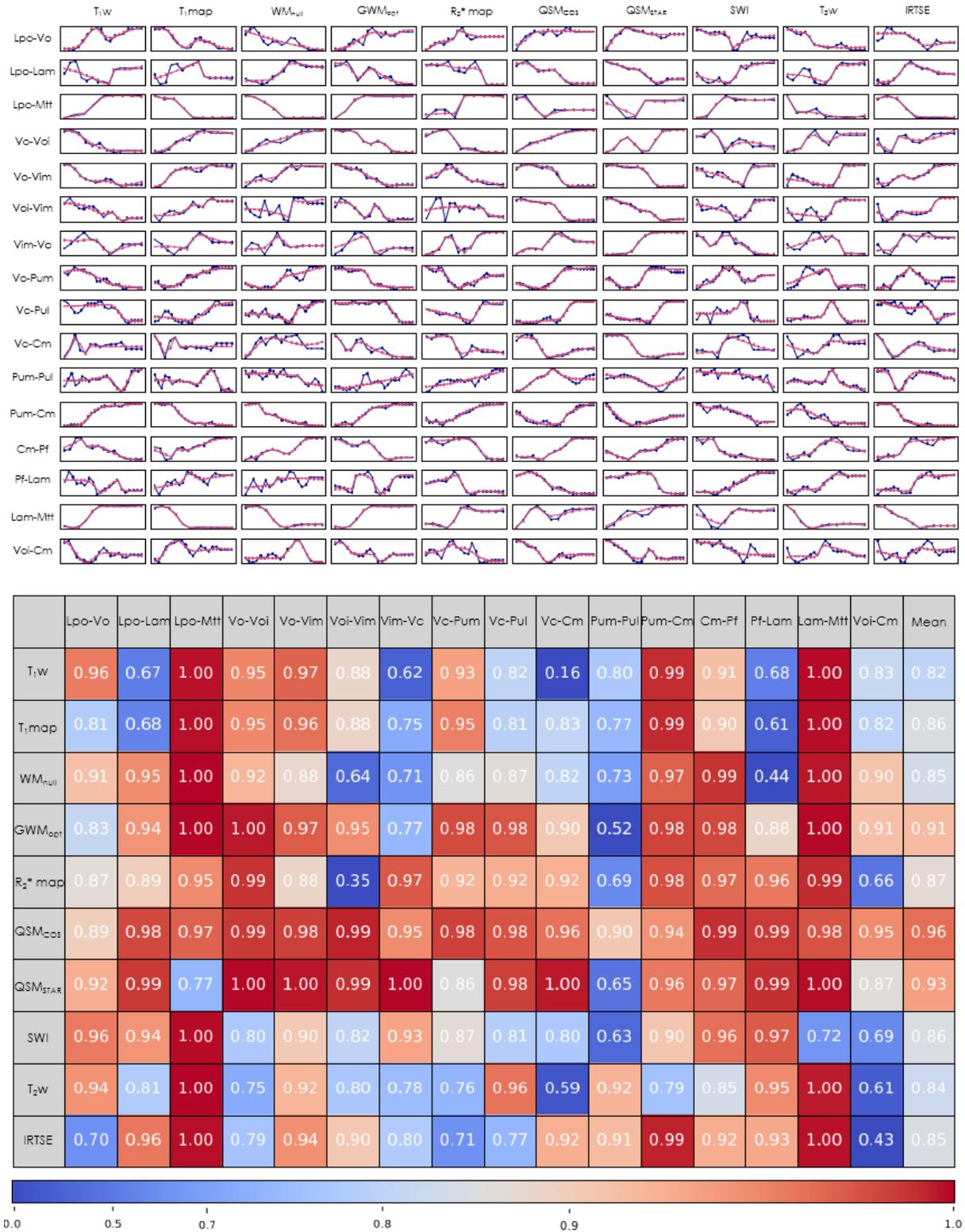

| | Lpo-Vo | Lpo-Lam | Lpo-Mtt | Vo-Vol | Vo-Vim | Vol-Vim | Vim-Vc | Vc-Pum | Vc-Pul | Vc-Cm | Pum-Pul | Pum-Cm | Cm-Pf | Pf-Lam | Lam-Mtt | Vol-Cm | Mean |
|---|---|---|---|---|---|---|---|---|---|---|---|---|---|---|---|---|---|
| $T_1$w | 0.96 | 0.67 | 1.00 | 0.95 | 0.97 | 0.88 | 0.62 | 0.93 | 0.82 | 0.16 | 0.80 | 0.99 | 0.91 | 0.68 | 1.00 | 0.83 | 0.82 |
| $T_1$map | 0.81 | 0.68 | 1.00 | 0.95 | 0.96 | 0.88 | 0.75 | 0.95 | 0.81 | 0.83 | 0.77 | 0.99 | 0.90 | 0.61 | 1.00 | 0.82 | 0.86 |
| WM$_{null}$ | 0.91 | 0.95 | 1.00 | 0.92 | 0.88 | 0.64 | 0.71 | 0.86 | 0.87 | 0.82 | 0.73 | 0.97 | 0.99 | 0.44 | 1.00 | 0.90 | 0.85 |
| GWM$_{sgr}$ | 0.83 | 0.94 | 1.00 | 1.00 | 0.97 | 0.95 | 0.77 | 0.98 | 0.98 | 0.90 | 0.52 | 0.98 | 0.98 | 0.88 | 1.00 | 0.91 | 0.91 |
| $R_2^*$ map | 0.87 | 0.89 | 0.95 | 0.99 | 0.88 | 0.35 | 0.97 | 0.92 | 0.92 | 0.92 | 0.69 | 0.98 | 0.97 | 0.96 | 0.99 | 0.66 | 0.87 |
| QSM$_{cos}$ | 0.89 | 0.98 | 0.97 | 0.99 | 0.98 | 0.99 | 0.95 | 0.98 | 0.98 | 0.96 | 0.90 | 0.94 | 0.99 | 0.99 | 0.98 | 0.95 | 0.96 |
| QSM$_{sp+k}$ | 0.92 | 0.99 | 0.77 | 1.00 | 1.00 | 0.99 | 1.00 | 0.86 | 0.98 | 1.00 | 0.65 | 0.96 | 0.97 | 0.99 | 1.00 | 0.87 | 0.93 |
| SWI | 0.96 | 0.94 | 1.00 | 0.80 | 0.90 | 0.82 | 0.93 | 0.87 | 0.81 | 0.80 | 0.63 | 0.90 | 0.96 | 0.97 | 0.72 | 0.69 | 0.86 |
| $T_2$w | 0.94 | 0.81 | 1.00 | 0.75 | 0.92 | 0.80 | 0.78 | 0.76 | 0.96 | 0.59 | 0.92 | 0.79 | 0.85 | 0.95 | 1.00 | 0.61 | 0.84 |
| IRTSE | 0.70 | 0.96 | 1.00 | 0.79 | 0.94 | 0.90 | 0.80 | 0.71 | 0.77 | 0.92 | 0.91 | 0.99 | 0.92 | 0.93 | 1.00 | 0.43 | 0.85 |

0.0   0.5   0.7   0.8   0.9   1.0

Supp. Fig. 7. Top: Intensity profiles of nuclei pairs for subject 3, with reference to the Schaltenbrand atlas at slice z = 3.5mm, for the different MRI contrasts. Each graph shows the intensity profile (blue), and the corresponding model fit (pink). Bottom: $R^2$ values obtained from the profile model fits of each pair of nuclei and contrast for subject 3. The mean $R^2$ value across all nuclei pairs for each contrast is displayed on the right-most column. A matching color (in logarithmic scale) was added behind each corresponding value, to aid visualization.

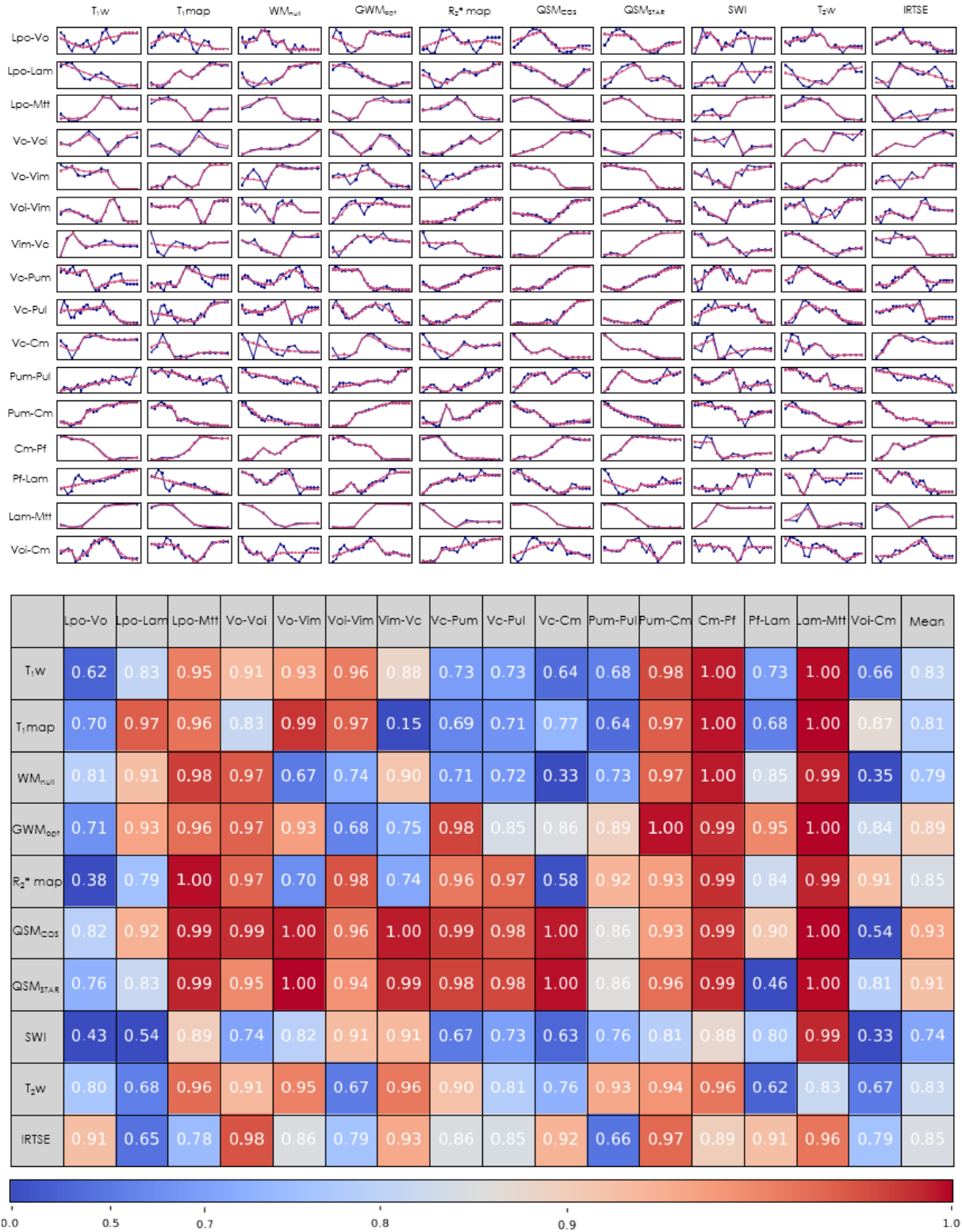

Supp. Fig. 8. Top: Intensity profiles of nuclei pairs for subject 4, with reference to the Schaltenbrand atlas at slice z = 3.5mm, for the different MRI contrasts. Each graph shows the intensity profile (blue), and the corresponding model fit (pink). Bottom: $R^2$ values obtained from the profile model fits of each pair of nuclei and contrast for subject 4. The mean $R^2$ value across all nuclei pairs for each contrast is displayed on the right-most column. A matching color (in logarithmic scale) was added behind each corresponding value, to aid visualization.

# Supplementary material

Snippet of MATLAB code publicly available in *https://github.com/JosePMarques/MP2RAGE-related-scripts*, demonstrating the extraction of an $R_1$ map from an MP2RAGE image, applicable for customized contrast variations of the sequence.

```matlab
MP2RAGEimg = load_untouch_nii(fullfile(phantom_path,'uni.nii.gz'));
INV1 = load_untouch_nii(('inv1.nii.gz'));
INV2 = load_untouch_nii(('inv2.nii.gz'));
MP2RAGEimg.img = double( MP2RAGEimg.img );
INV1.img = double( INV1.img );
INV2.img = double( INV2.img );

% First we correct the INV1 and INV2 maps, to be with negative and positive value
[INV1final, INV2final] = Correct_INV1INV2_withMP2RAGEuni(INV1,INV2,MP2RAGEimg,[]);
INV1img = INV1final.img;
INV2img = INV2final.img;

B1 = load_untouch_nii(('B1_map.nii.gz'));

% Then we create the fingerprint for T1 and PD
[T1, PD, R1] = MP2RAGE_dictionaryMatching(MP2RAGE,INV1img,INV2img,B1.img);
```